\documentclass[fleqn,usenatbib]{mnras}

\usepackage{newtxtext,newtxmath}

\usepackage[T1]{fontenc}
\DeclareRobustCommand{\VAN}[3]{#2}
\let\VANthebibliography\thebibliography
\def\thebibliography{\DeclareRobustCommand{\VAN}[3]{##3}\VANthebibliography}

\usepackage{float}

\usepackage{graphicx}	
\usepackage{amsmath}	
\usepackage{comment}

\title[Hydrodynamic Natal Kicks and Formation of Black Hole Binaries]{Hydrodynamic Natal Kicks and Formation of Black Hole Binaries}

\author[Santiago-Leandro, Avilés-Niebla \& Batta]{
Edwin Santiago-Leandro,$^{1,2}$\thanks{E-mail: esantiago@inaoep.mx}
Clarissa Avilés-Niebla,$^{3}$
and Aldo Batta$^{2}$
\\
$^{1}$Universidad de Costa Rica, San José, San Pedro, 11501, Costa Rica\\
$^{2}$ Instituto Nacional de Astrofísica Óptica y Electrónica, Luis Enrique Erro 1, Tonantzintla CP 72840, Puebla, México\\
$^{3}$Universidad de Guanajuato, Guanajuato, México
}

\date{Accepted XXX. Received YYY; in original form ZZZ}

\pubyear{\the\year{}}

\begin{document}
\label{firstpage}
\pagerange{\pageref{firstpage}--\pageref{lastpage}}
\maketitle

\begin{abstract}
The observation of over 300 Gravitational Wave (GW) events by the LIGO-Virgo-KAGRA collaboration, has yielded information of the merger of compact object binaries, such as their individual masses and spins. However, the effective spin parameter obtained from the GW signal provides important information that could be used to discern the binary's formation channel. Overall, binaries formed from dynamic interactions in dense stellar environments are expected to have a random distribution of effective spins between -1 and 1. Meanwhile, binaries formed from the evolution of isolated stellar binaries are expected to have positive effective spins unless there is a large enough natal kick during the compact object formation that could significantly alter the orientation of the individual spins with respect to the binary's angular momentum. We study the viability of obtaining large natal kicks during the formation of massive Black Holes by studying the asymmetric mild explosion of $13\text{M}_{\odot}$, $21\text{M}_{\odot}$, $28\text{M}_{\odot}$ and $34\text{M}_{\odot}$ pre-supernova stellar models, by means of 3D Smooth Particle Hydrodynamics simulations. Our results show that, even kicks in the order of $\sim$ 1000 km/s, which are among the highest values observed for natal kicks, can't consistently change the orientation of the orbit. Physical setups where the orientation of the orbit changes require the kick to be directed correctly, and they would also require for the supernova explosion energy to be quite high.
\end{abstract}




\section{Introduction}

Since the first discovery of Gravitational Waves (GWs) in 2015, a new source of information was available for the study of compact binary systems composed by black holes (BHs), neutron stars (NSs) or both. The more than 90 GWs events detected by the LIGO-Virgo-KAGRA (LVK) collaboration up until the O3 observing run allow to study the properties and formation channels of binary black holes (BBHs) \citep{abbott}. From the GW detections, it is possible to discern the individual masses of the black holes (BHs), as well as their effective spin parameter 

\begin{equation}
    \chi_{\text{eff}} = \left( \frac{M_{\rm bh,1}\vec{\boldsymbol{S}}_{1}+M_{\rm bh,2}\vec{\boldsymbol{S}}_{2}}{M_{\rm bh,1}+M_{\rm bh,2}} \right) \cdot \hat{\boldsymbol{L}},
    \label{chi}
\end{equation}

\noindent where the index $i=1,2$ refers to the primary or secondary BH respectively, $\vec{\boldsymbol{S}}_{i} = c\vec{\boldsymbol{J}}_{i}/(GM_{\rm bh,i}^2)$ is the dimensionless spin parameter, $\vec{\boldsymbol{J}}_{i}$ is the angular momentum of the BH with mass $M_{\rm bh,i}$ and $\hat{\boldsymbol{L}}$ is a unitary vector in the direction of the orbital angular momentum. This quantity ranges from -1 to 1, depending on how aligned the individual spins of the BHs are with respect to the orientation of the orbit \citep{ligo-virgo,rodriguez}. Depending on the formation channel of the BBH, different possible distributions of $\chi_{\text{eff}}$ are expected \citep{rodriguez}. 

For BBHs that are created from dynamical interactions in dense stellar environments, we expect a random distribution of $\chi_{\text{eff}}$ between -1 and 1. This is due to the numerous interactions the BBHs would experience in this environment \citep{benacquista, coleman, mapelli}. In general, these systems need a mechanism for the BHs to be close enough so that they would merge in a timescale shorter than the age of the Universe. In the case of the BBHs from dense stellar environments, we expect single-binary, binary-binary, and even single-single interactions to help the binary system loose orbital energy, and develop a compact binary as a result. There is no reason to think that these types of interactions would produce BBHs with spins aligned with the angular orbital momentum of the system \citep{coleman}.

Another mechanism for the formation of BBHs would be isolated stellar binary systems. These systems begin as a stellar binary system of two OB type stars, which interact with one another throughout their evolution, whenever one of them fills its respective Roche Lobe. The mechanism at play responsible of decreasing the orbital separation of the system is a Common Envelope (CE) that forms when the second star fills its Roche Lobe, which would reduce the orbital separation to the order of $\sim 10 \text{R}_{\odot}$ \citep{belczynski, bavera}. From these systems, one would expect the individual spins of the binary components to be aligned with the orbital angular momentum. This would mean that, from isolated stellar binary systems, one would initially expect a positive distribution of values for $\chi_{\text{eff}}$, unless another mechanism is at play \citep{mapelli, rodriguez}. For instance, \cite{baibhav} analyzed a combination of other possible mechanisms and found a viability of isolated stellar systems to have misaligned spins. It is also possible that an asymmetric fallback of material may be able to change the direction of the BH spin, particularly for lower mass BHs \citep{janka2022}.

Another possibility is that a BH natal kick (NK) of linear momentum can be theoretically able to change the direction of the orbit, and as a result, the sign of $\chi_{\text{eff}}$. This NK can be the result of an asymmetry in the Supernova Explosion (SNE) \citep{janka2013, brandt}. If the integrated momentum of the SNE tends towards a particular direction, the compact object would go to the opposite direction to conserve linear momentum. Natal kicks have been observed numerous times in Neutron Stars (NS), but their detection in BHs is still limited \citep{repetto}. 

Theoretical calculations have been made to determine how big the NK should be to change the sign of $\chi_{\text{eff}}$ \citep{brandt, janka2013}, and more recently this analysis shows that it is unlikely that a BH NK could by itself be the mechanism at play \citep{stegmann}. In the simplest scenario one would expect that the NK received by the NS that eventually collapses into a BH would be reduced by the accretion process. However, \cite{janka2013} proposed that the BH NK would be larger, by means of a gravitational tug-boat mechanism. In this work, it was proposed that the asymmetric accretion and distribution of momentum around the accreting NS, could lead to a larger final velocity $v_{\rm {BH,f}}$ than the one expected after a spherically symmetric accretion of mass and momentum $v_{\text{sym}}$ according to:

\begin{equation}
    v_{\rm {BH,f}} > v_{\text{sym}} \equiv \frac{M_{\text{NS}}}{M_{\text{BH,f}}}v_{\text{NS}},
    \label{vk_sym}
\end{equation}

\noindent where $v_{\text{BH,f}}$ is the final BH velocity, $M_{\text{BH,f}}$ is the final BH mass and $M_{\text{NS}}$ and $v_{\text{NS}}$ are the initial NS mass and velocity respectively. How strong this mechanism may be could also depend on the type of SNE \citep{janka2017}. Recent work has also shown that it is not necessarily the case that there will be an alignment between the orientation of the NK and the compact object spin \citep{janka2022}.

Analytic calculations from \eqref{vk_sym} provide a good first-order approximation to the expected magnitude of the BH NKs that will prevail after the recently formed NS substantially increases its mass and turns into a BH after accreting the collapsing envelopes outside the stellar core. However, these calculations do not take into account any asymmetry in the accretion process, which would yield to momentum being accreted asymmetrically and ultimately to a different reduction to the natal kick. Thus, this work will focus on studying the accretion process of asymmetric explosions by a newly formed BH located at the center of an exploding star, whose NK is  determined by the asymmetry and energy of the explosion. This will be done by means of 3D Smooth Particle Hydrodynamics (SPH) simulations of pre-supernova stellar models with different masses, which will allow us to follow the accretion process and the entire evolution of the BH's properties (including the NK). Once we determine if the reduced kick velocity is larger as \cite{janka2013} proposed, we will make use of our results to explore scenarios where the kicks are large enough to invert the binary's angular momentum, leading to a change in sign of $\chi_{\text{eff}}$.


\section{Physical Setup and Numerical Methods}

\label{ph-setup}

As stated before, by performing 3D SPH simulations of asymmetric explosions, we will study the accretion process and the natal kicks induced on a newly formed BH located at the center of an exploding pre-supernova star. We will make use of a modified version of the code GADGET (\textbf{GA}laxies with \textbf{D}ark matter and \textbf{G}as int\textbf{E}rac\textbf{T}), that has been used in similar previous works \citep{batta-ramirez,schrøder, Vigna_2021}. 

In order to be able to discern the effect of asymmetric "failed" explosions on the BH's natal kick velocity $v_{\rm BH,i}$, from the effects caused by other physical processes, we focused primarily on simulations of isolated stars. This will ensure that the accretion process is not affected by the companion, which for small enough orbital separations can induce a large torque on infalling material, leading to changes in the accretion process and ultimately, in the BH's spin \citep{batta-ramirez,schrøder}. Therefore, in the first part of this work, we will focus on simulations of isolated stars in order to study the effect of asymmetric explosions and the subsequent accretion fallback on the BH's natal kick and overall properties. Then, we will compare a subset of the results from isolated stars, to results from simulations including a binary companion. Below we will describe the physical model and how the initial conditions are constructed.


\subsection{Initial Setup for Exploding Isolated Stars}
\label{sec:Init_singles}

In order to build the initial conditions of our simulations, we made use of four pre-supernova stellar models from \cite{heger} with masses ranging from $13\text{M}_{\odot}$ to $34\text{M}_{\odot}$, which are low metallicity, rotating He stars without H envelope, similar to stars obtained after multiple mass transfer episodes in the isolated binaries formation channel \citep{belczynski}. The models are mapped into a 3D distribution of SPH particles reproducing the collapsing star's mass profile and hydrodynamical properties. This mapping is done by distributing SPH particles on spherical shells according to the discrete surface mapping used in HEALPix \citep{gorski}. The number of SPH particles $N_{\rm shell}$ at a radius $r$, is given by $\rho(r)$ and the particles' mass, and the spherical shells are randomly rotated to avoid any artificial alignment in the particle distribution. The interested reader can find more details about this setup in \cite{batta-ramirez}, based on the initial setup from \cite{Pakmor2012}.

Following \cite{janka2013}, we assume that the BH's natal kick originates from an asymmetric emission of neutrinos during the stage when the core is still a NS and has not collapsed into a BH. We also assume that this asymmetric emission of neutrinos (and any other explosion mechanism present) leads to an asymmetric injection of kinetic energy into the layers surrounding the stellar core which yields a mild (or "failed") explosion. This asymmetric explosion will determine the intensity and direction of the kick imparted initially to the recently formed BH. In order to avoid dealing with the complicated accretion process onto a NS, we assumed that the innermost $3M_{\odot}$ of the star has already collapsed into a BH right after producing this asymmetric mild explosion onto the surrounding layers of the star.

\subsubsection{Asymmetric explosion and natal kick}\label{AsymmetricICs}

\begin{figure}
    \centering
\includegraphics[trim={2.0cm 3cm 3.0cm 2.5cm},width=\columnwidth]{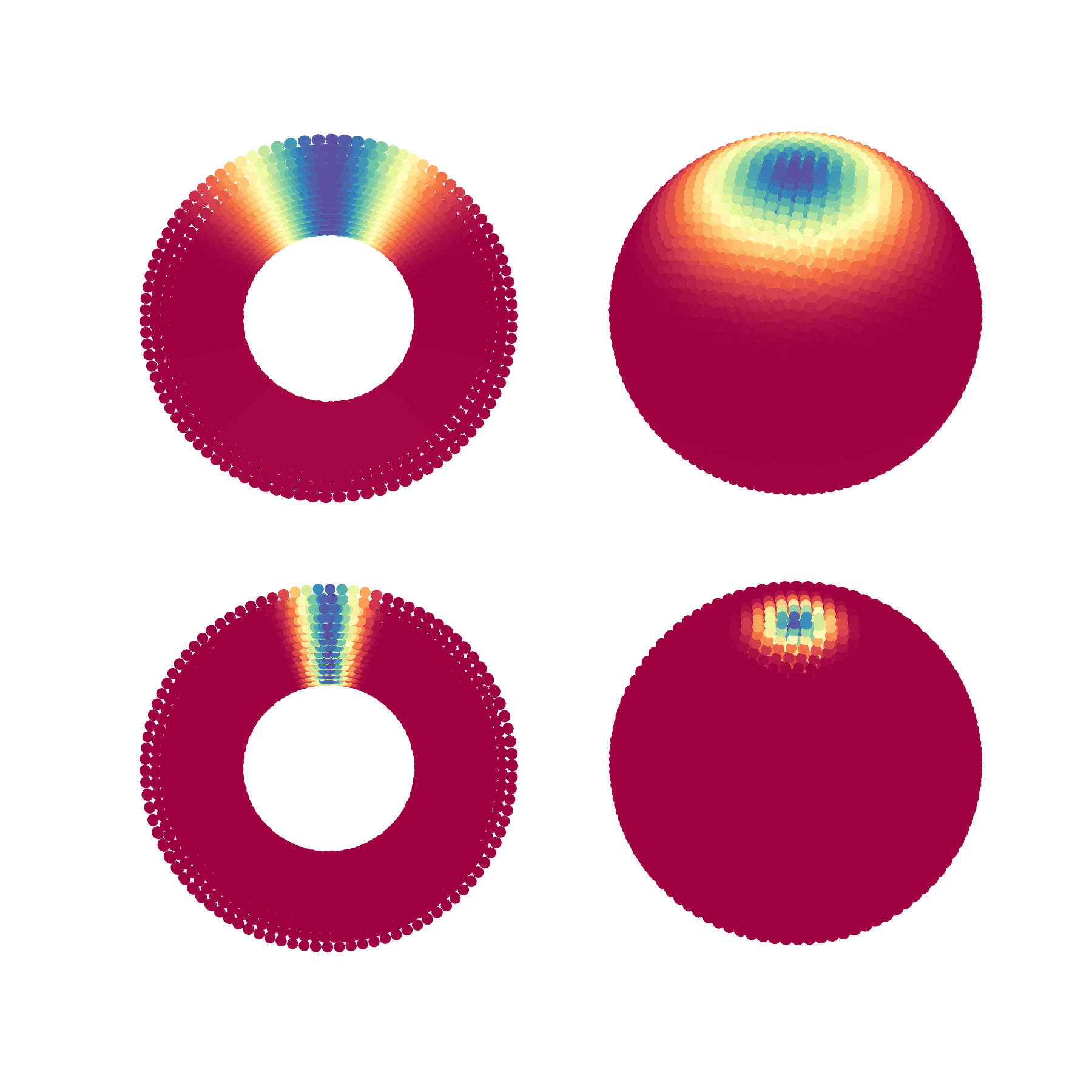}

\includegraphics[trim={1cm 0.0cm 1.0cm 0cm},width=\columnwidth]{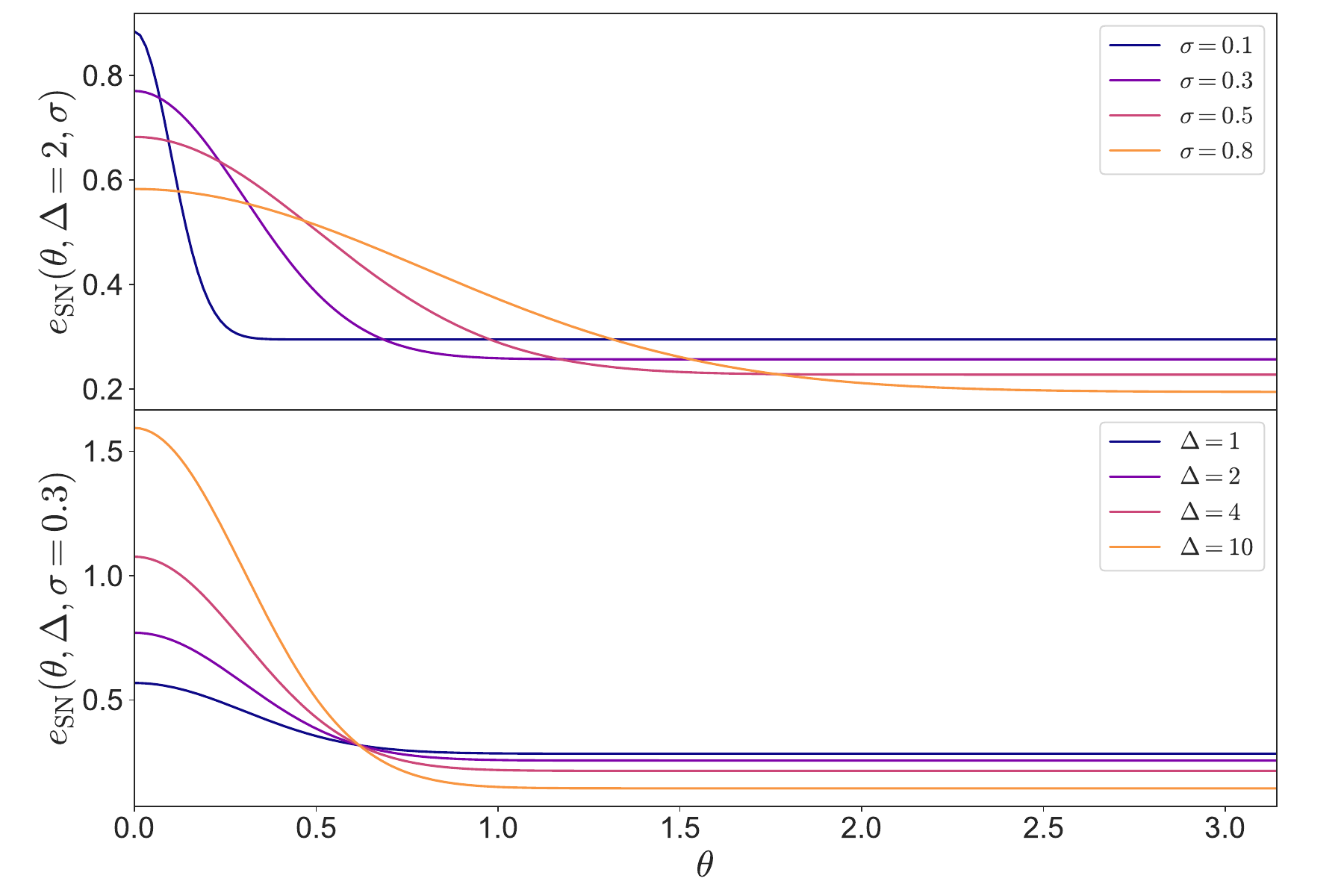}

    \caption{Bottom panels: Normalized energy distribution $e_{\rm{SN}}(\theta,\Delta,\sigma)$ given to SPH particles surrounding the BH, for different values of $\Delta$ (bottom) and $\sigma$ (top). Top panel: Distribution of kinetic energy (radial velocity) on particles surrounding the BH for $\sigma=0.3$ (top) and $\sigma=0.1$ (bottom), for $\Delta=10$. The largest radial velocities (blue) are located on particles along the direction of the asymmetry (the z-axis in this example) and decreases for particles at larger polar angles until reaching a floor value.}
    \label{vprofile}
\end{figure}

To simulate the asymmetric explosion, kinetic energy was injected into the SPH particles within the innermost $1M_\odot$ surrounding the black hole. We treated the explosion energy, $E_{\rm SN}$, as a free parameter scaled relative to the binding energy of the envelope surrounding the BH, $E_{\rm bn}$. By constraining the ratio to $0 \leq E_{\rm SN}/E_{\rm bn} \leq 1$, we ensure that the explosion energy does not exceed the threshold required to unbind the rest of the star.
This energy is distributed according to a Gaussian function of the polar angle, $\theta \in [0, \pi]$:

\begin{equation}
    f(\theta,\Delta,\sigma) = 1 + \Delta e^{-\theta^2/2\sigma^2},
\end{equation}
where, $\Delta$ is a parameter defining the amplitude of the energy difference between the direction of the asymmetry ($\theta=0$) and far from it (towards $\theta=\pi$). Thus, once the parameter $E_{\rm{SN}}$ is defined, the explosion energy per unit mass given to an SPH particle $i$ located at a polar angle $\theta_i$ is given by:
\begin{equation}
    e_{\rm{SN},i}(\theta_i,\Delta,\sigma)= \dfrac{f(\theta_i,\Delta,\sigma)}{\sum_i f(\theta_i,\Delta,\sigma) m_i} E_{\rm SN},
    \label{eSN_part}
\end{equation}
where the sum is done over all particles accounting for the innermost $1M_{\odot}$. This guarantees that the explosion energy $E_{\rm SN}$ is recovered when adding the explosion energy given to each SPH particle, and allows us to control the intensity and shape of the asymmetry (and the kick) using the parameters $\Delta$ and $\sigma$. The bottom panels from Figure \ref{vprofile}, show how the energy distribution given to the SPH particles $e_{\rm{SN}}(\theta,\Delta,\sigma)$ is affected by $\Delta$ and $\sigma$ when maintaining the same explosion energy $E_{\rm SN}$ (normalized in the plot). Meanwhile, the top panel from Figure \ref{vprofile} shows how this gaussian energy distribution translates into the distribution of radial velocities of the particles ranging from the maximum values at $\theta=0$ to the minimum values away from the asymmetry.

The overall set of parameters utilized for the simulations can be found in Table 
\ref{Tab:SNEfracs}. The range of natal kicks $v_{\rm kick}$ in our simulations was chosen in order to explore velocities which, when directed in the exact opposite direction of the orbital motion, are able to invert the direction of the binary's orbital angular momentum, and therefore, its effective spin $\chi_{\rm eff}$. Thus, the kick velocities must be comparable or larger than the binary's orbital velocity $v_{\rm orb}$, which can range from a few hundreds to a few thousands of km/s, depending on the masses and separation of the components. However, considering the natal kick velocity distributions observed for neutron stars \citep{verbunt} and BHs \citep{repetto,mandel2016}, kick velocities rarely exceed $v_{\rm kick}\approx 1000$ km/s. Consequently, our assumption of $v_{\rm kick} = 1500$ km/s represents an extreme, physically unlikely scenario. This value is explored primarily as a limiting case to determine the kick magnitude needed to produce negative  $\chi_{\rm eff}$ in specific systems.

The asymmetry was set up by fixing $\sigma = 0.6$, and varying $\Delta$ so it has a value such that it would correspond to the different NKs mentioned in Table \ref{Tab:SNEfracs} (which corresponds to  $\Delta \in [0.25,400]$). It is worth mentioning that for the $13M_{\odot}$ preSN star, there is no simulation with explosion energy of $E_{\rm SN}= 0.4E_{\rm bin}$ and natal kick of 1500 km/s, and neither simulations with $E_{\rm SN}= 0.2E_{\rm bin}$ and natal kicks of 1000 km/s or 1500 km/s. This is due to the fact that this particular stellar model has the lowest binding energy, which determines the available kinetic energy in the explosion to obtain a certain natal kick. For the energies and kick velocities mentioned before, there were no values for $\Delta$ that produced the required natal kicks. The rest of the stellar models have the complete set of parameters explored.

\subsubsection{Accretion and BH properties}

As mentioned before, we assume that the collapse of the innermost stellar material has already produced a BH with mass $M_{\rm bh,2,i} = 3M_{\odot}$, which is treated as a sink particle initially located at the origin. The BH will acquire a natal kick according to asymmetries in the SPH particle distribution around it and to the asymmetry of the explosion. Firstly, since the SPH particle distribution is never exactly spherically symmetric, there will be a slight asymmetry in the velocity if rotation or a symmetrical explosion (radial velocity) is added to the SPH particles. Depending on the simulation's resolution, these small departures from spherical symmetry induce a natal BH kick of a few km/s, negligible for the kick intensities explored in our work (Table \ref{Tab:SNEfracs}). 

As in previous works \citep{batta-ramirez,schrøder,Vigna_2021}, the accretion of SPH particles by the BH (sink particle), is decided in terms of an accretion radius $r_{\rm acc}=0.01 R_{\odot}$ and the particle's angular momentum compared to the angular momentum of the innermost stable circular orbit (ISCO) around the BH with mass $M_{\rm BH}$ and spin $S_{\rm BH}$ (i.e when $j< j_{\rm isco}$). All of our simulations begin with a non rotating BH ($S_{\rm BH}=0$), whose properties will be updated as it accretes mass, and momentum (linear and angular) from the SPH particles. 

\begin{table}
\centering
\caption{Explosion energies (top row), kick velocities (middle row) and asymmetry width $\sigma$ (bottom row) used in our simulations.}
\begin{tabular}{c c c}
\hline
Parameter & \hspace{1cm} & Values explored \\
\hline
$E_{\rm SN}/E_{\rm bin}$ & \hspace{1cm} & 0.2 \hspace{0.3cm} 0.4 \hspace{0.3cm} 0.6 \hspace{0.3cm} 0.8 \hspace{0.3cm} 0.9 \\ 
\hline
$v_{\rm kick}$ (km/s) & \hspace{1cm} & 200 \hspace{0.3cm} 500 \hspace{0.3cm}  1000 \hspace{0.3cm}  1500  \\ \hline
 $\sigma$ & & 0.6 \\ \hline
 \end{tabular}
\label{Tab:SNEfracs}
\end{table}



\subsubsection{Stellar rotation and convergence tests}


For our simulations, the SPH particles conforming the star are given a constant angular velocity $\Omega$ around the $Z$-axis with a rotation period of $0.5$ days. Following \cite{Batta_2019}, we computed the spin of the BH formed from the complete collapse of the stellar models, obtaining values ranging from 0.025 to 0.125, and spins that range from 0.001 to 0.005 for the $3M_{\odot}$ BH produced from the innermost mass, allowing us to safely assume a spin of zero for the initial BH. Moreover, small rotations guarantee small deviations from spherical symmetry due to centrifugal forces, allowing us to isolate the effect of asymmetric explosions in the accretion process.

For the convergence tests we used resolutions ranging from 1 to 16 million SPH particles. While higher resolutions resulted in a slight decrease in the BH's final mass and minor variations in the kick velocity, the overall discrepancies in accretion rates, mass, and velocity remained within a few percent. Consequently we adopted a standard resolution of $10^6$ particles to efficiently explore the parameter space, reserving a higher resolution of $8 \times 10^6$ particles for the binary companion simulations discussed below.

All of our simulations were run for 2 solar dynamical timescales ($t_{\rm dyn,\odot} = (R_{\odot}^3/GM_{\odot})^{1/2}\simeq 1600$ s), providing enough time for most of the mildly ejected material to fall back onto the BH. The remaining amount of material that takes longer to fall back is negligible and will not be able to produce significant changes in the BH's properties.

\subsection{Initial setup for binary systems}
\label{initBinaries}

From the isolated star simulations we will obtain information about how asymmetric explosions and the subsequent accretion, dictate the final mass and velocity of the BH. However, we still need to determine whether these properties result in binary systems with negative effective spin parameters ($\chi_{\rm eff}$). This requires incorporating the orbital dynamics and the properties of the companion.



We defined the companion BH properties based on observations from \cite{abbott2023}, which indicate that most observed black hole binary (BHB) mass ratios are consistent with the equal-mass limit, $q = M_{\rm bh,2}/M_{\rm bh,1} \sim 1$. Since the pre-SN stars in our models can lose a significant fraction of their mass during the explosion, we selected companion masses slightly lower than the progenitor's pre-SN mass, ensuring the second-formed BH remains the less massive component and maintaining a mass ratio of $q \lesssim 1$. 

To maintain a tractable parameter space, we adopted a fixed pre-supernova mass ratio of $M_{\rm bh, 1}/M_{\rm preSN}=0.933$ for all simulations. Even though \cite{Qin} suggests that the first born BH formed from the isolated binary formation channel typically possesses a negligible spin, we assumed an initial spin parameter of $S_{\rm BH, 1}=0.2$ for the primary component. This assumption guarantees that the sum of the individual weighted spins of the BHs (shown in Eq. \ref{chi}) is always clearly distinguishable as positive, and that the effective spin $\chi_{\rm eff}$ can only become negative if the orbital angular momentum is inverted.



Regarding the binary’s orbital parameters, we assume the system originates from a Common Envelope (CE) phase \citep{Taam_2010,Ivanova_2013} eventually resulting in a close binary composed by a BH and a Helium star in an almost circular orbit. Consequently, we adopted circular orbits with separations slightly below the maximum threshold required for a gravitational-wave (GW) driven merger. However, we should keep in mind that the post CE eccentricity may not be negligible \citep{Glanz_2021}, and that circularization processes after the CE phase may not be effective in circularizing the orbit. The specific properties used in our simulations are summarized in Table \ref{Tab:Bin}, including the orbital velocities for each binary, which already informs us that in order to invert the orbital direction of the recently formed BH, it must receive a kick velocity above $\sim 500$ km/s.

\begin{table}
\centering
\caption{Orbital parameters taken for each pre-supernova stellar model. All orbits are assumed to be initially circular.}
\begin{tabular}{|l|l|l|c|}
\hline
$M_{\rm preSN}$              & $M_{\rm bh,1}$                 & $a_{0}$  & $v_{\rm orb}$           \\ \hline
$13M_{\odot}$ & $13.93M_{\odot}$ & $25R_{\odot}$ & 453.33 km/s\\ \hline
$21M_{\odot}$ & $22.5M_{\odot}$  & $35R_{\odot}$  & 486.94 km/s\\ \hline
$28M_{\odot}$ & $30M_{\odot}$ & $40R_{\odot}$  &665.29 km/s\\ \hline
$34M_{\odot}$ & $36.43M_{\odot}$    & $54R_{\odot}$  & 733.12 km/s\\ \hline
\end{tabular}
\label{Tab:Bin}
\end{table}

\subsubsection{Instantaneous Mass Loss Approximation}
\label{IML}
Since the SNe occurs in a dynamical timescale, which is much smaller than the orbital period of the system, it is usually safe to assume that the SNe, the NK, and the changes in the system they produce, are instantaneous. This so-called \textit{instantaneous mass loss approximation} (IML) allow us to obtain analytically, the orbital properties of the binary after one of the components loses mass and receives a natal kick, due to a SN explosion \citep{brandt}. Using this approximation we can obtain binary properties for the simulations of single stars described in \ref{sec:Init_singles}, given the orbital properties of the corresponding companion shown in Table \ref{Tab:Bin}.

Thus, for a circular binary system with orbital separation $a_0$, and masses $M_{\rm preSN}$ and $M_{\rm bh,1}$, the SN explosion causes an instantaneous mass loss from $M_{\rm preSN}$ and the system, producing a remnant of mass $M_{\rm bh,2}$. Defining $\psi$ as the angle between the kick velocity $\vec{v}_{\rm kick}$ and the orbital plane, and $\alpha$ as the angle between the projection of $\vec{v}_{\rm kick}$ on the orbital plane and the orbital velocity $\vec{v}_{\rm orb}$, one can write the orbital energy $E_{\rm IML}$, separation $a_{\rm IML}$, and eccentricity $e_{\rm IML}$ of the binary after the instantaneous mass loss as:

\begin{align}
    E_{\rm IML} =&  -\frac{GM_{\rm bh,2}M_{\rm bh,1}}{a_0}\left[2-\tilde{m}\left(1+2\tilde{v}\cos{\alpha}\cos{\psi}+\tilde{v}^2\right)\right],
    \\
    \nonumber\\
    \frac{a_0}{a_{\rm IML}} =& \ 2-\tilde{m}[1+2\tilde{v}\cos{\alpha}\cos{\psi}+\tilde{v}^2],
    \label{afIML}
    \\
    \nonumber\\
    e^2_{\rm IML} =& \ 1-\tilde{m}\left[2-\tilde{m}\left(1+2\tilde{v}\cos{\alpha}\cos{\psi}+\tilde{v}^2\right)\right] \times\\ 
    &\left[(1+\tilde{v}\cos{\alpha}\cos{\psi})^2+(\tilde{v}\sin{\psi})^2\right],\nonumber
\end{align}
where $\tilde{m} = (M_{\rm bh,1}+M_{\rm bh,2})/(M_{\rm bh,1}+M_{\rm preSN})$ and $\tilde{v} = v_{\text{kick}}/v_{\text{orb}}$. Another important orbital property that can be obtained is the binary's orbital angular momentum after the explosion $L_{z,\rm IML}$:
\begin{equation}
    L_{z,\rm IML}= \frac{M_{\rm bh,2}M_{\rm bh,1}}{M_{\rm bh,2}+M_{\rm bh,1}}(v_{\text{orb}}+v_{\text{kick}}\cos{\psi}\cos{\alpha})\ a_0,
    \label{Lz}
\end{equation}

which can be used to calculate the effective spin $\chi_{\rm eff}$ by means of \eqref{chi}. We will define the original orbit as going counterclockwise, so that a change in the direction of the orbit will be seen as a negative $L_{z,\rm IML}$.

\subsubsection{Binary simulations}
\label{BinarySims}

Besides the binary properties obtained from the IML approximation, we ran additional simulations of the binary systems in Table \ref{Tab:Bin} including both the exploding star and the companion BH. Using a higher resolution of $8\times 10^{6}$ SPH particles and an natal kick velocity of 1000 km/s, we ran 5 simulations corresponding to the explosion energies shown in Table \ref{Tab:SNEfracs} for each one of the binaries in Table \ref{Tab:Bin}. The initially high natal kick will be significantly reduced by accretion according to \eqref{vk_sym}, ultimately determined by the explosion energy $E_{\rm SN}/E_{\rm bin}$. Thus, we expect to obtain kick velocities larger than the orbital velocities in Table \ref{Tab:Bin} for some of the larger energy explosions that lead to less fallback. These results will allow us to account for the gravitational pull from the second BH in the dynamics of the ejecta and the accretion on the recently formed BH, and to assess if the IML approximation provides a good estimate for the effective spin of the binary in such scenarios.

\section{Results}

\subsection{Tug-boat Mechanism in Exploding Isolated Stars}
\label{isolated}

\begin{figure}
  \centering
  \includegraphics[width=0.5\textwidth]{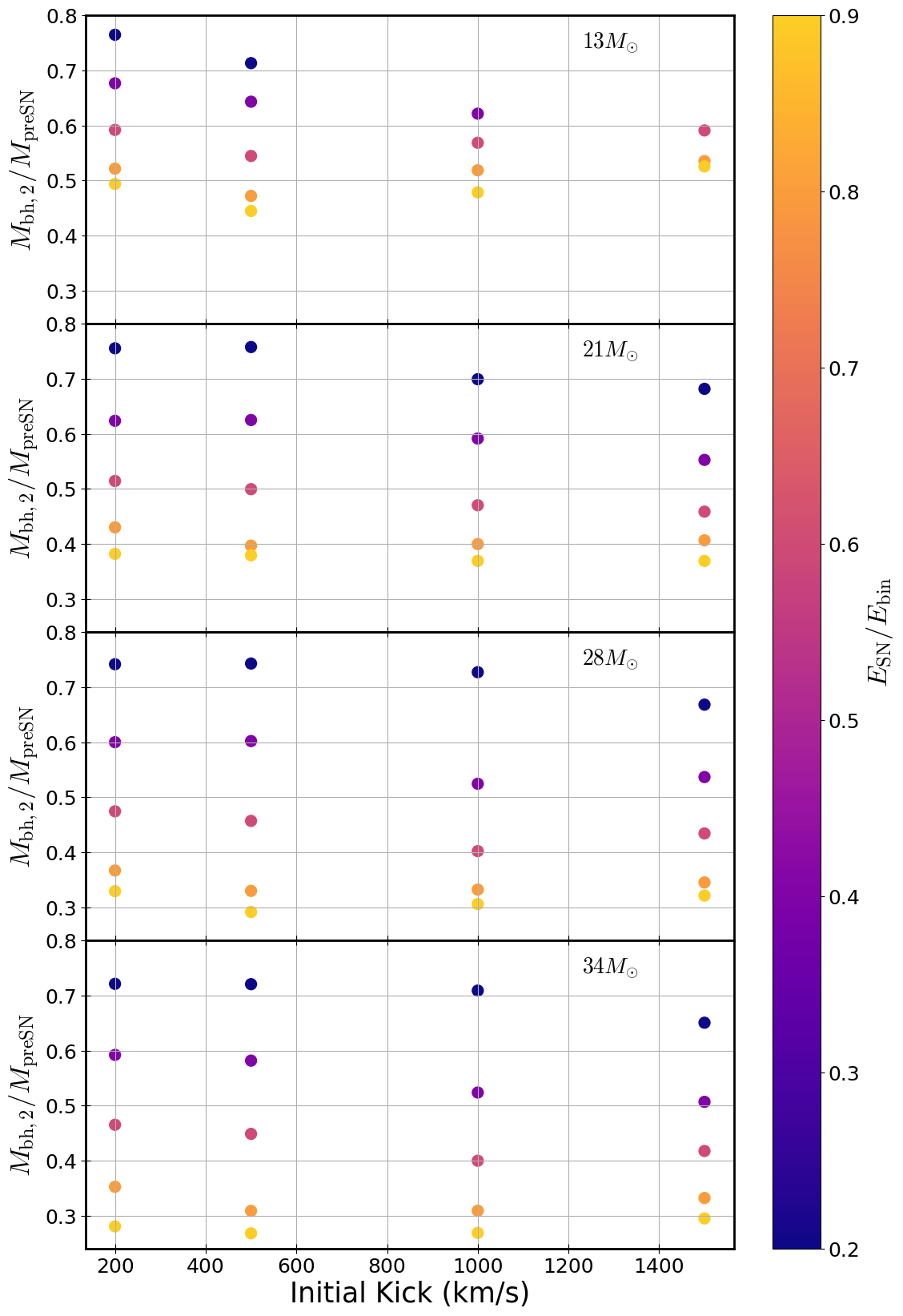} 
  \caption{Final BH mass at the end of the simulations (as a fraction of the initial star's mass) vs the initial BH kick velocity. Each panel corresponds to a different stellar model and the color of the points represents the explosion energy scaled by the envelope's binding energy $E_{\rm SN}/E_{\rm bin}$. The specific values utilized are in Table. \ref{Tab:SNEfracs}.}
  \label{BHMass}
\end{figure}

We begin by showing the results from the simulations with an isolated BH in the range of SNE energies and natal kicks mentioned in Table \ref{Tab:SNEfracs}.In Figure \ref{BHMass} we show the final BH mass at the end of the simulations as a function of the natal kick velocity. Each panel corresponds to a different preSN star and the color of the points indicates the explosion energy used in the simulation.

As expected, for a given natal kick velocity, a lower energy SNE is able to eject less material, leading to a more massive BH. However, the natal kick has a non trivial effect on the accretion flow, since changing the kick from 200 km/s to 500 km/s, leads to BH masses that stay similar or slightly decrease for all explosion energies. Meanwhile, increasing the kick from 500 km/s (or 1000 km/s) can lead to a noticeable increase or decrease on the BH's mass depending on the explosion energy, which is most obvious in the $13M_{\odot}$, $28M_{\odot}$ and $34M_{\odot}$ stellar models. 



\begin{figure}
  \includegraphics[width=0.5\textwidth]{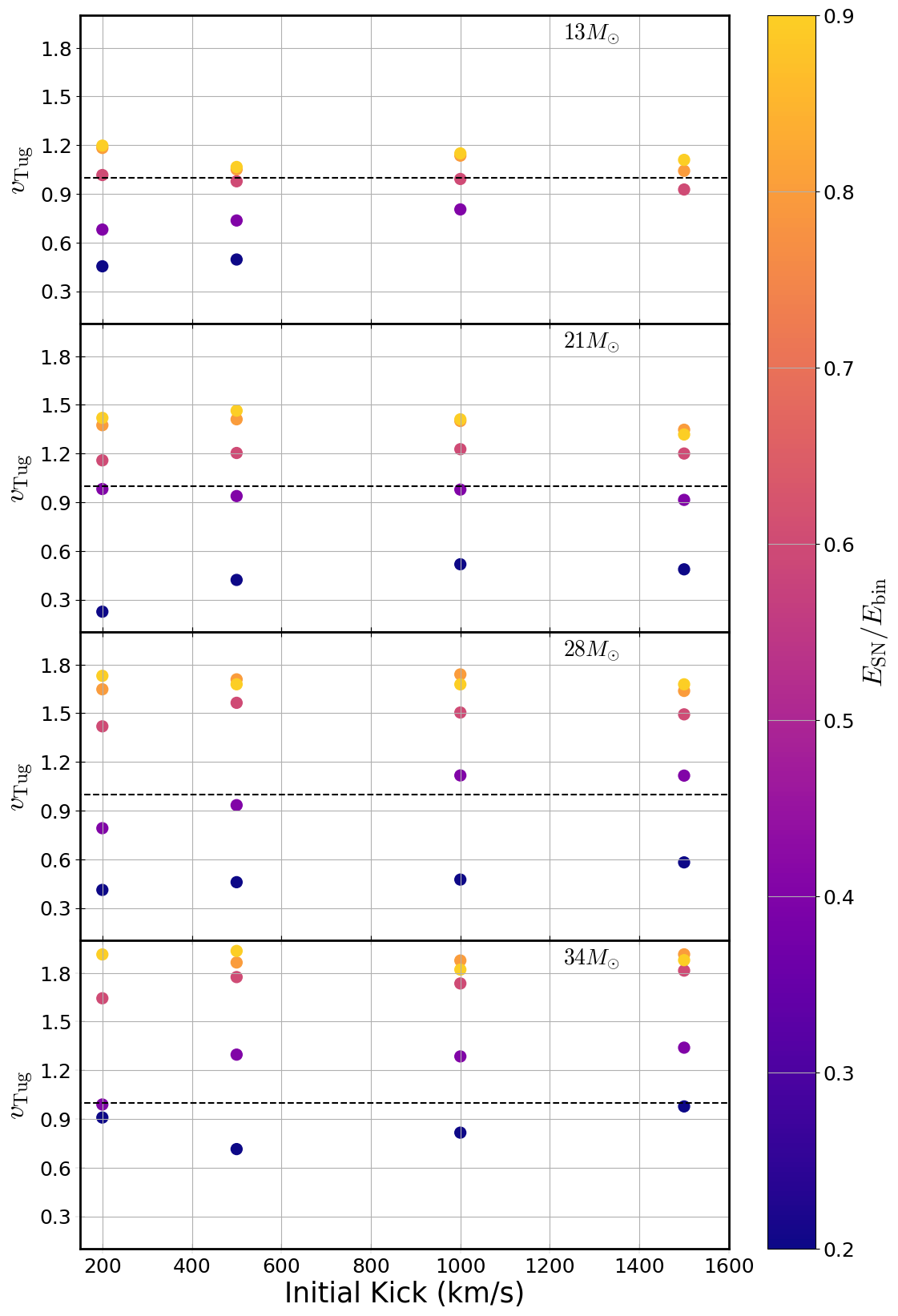}
  \caption{Final Tug-boat parameter $v_{\rm Tug}$ vs BH natal kick for all simulations. The horizontal dashed line indicates $v_{\rm Tug}=1$ where the final kick velocity from the simulations is the same as the kick velocity expected from symmetrical accretion using the final masses from Figure {\ref{BHMass}}. }
  \label{vfracs}
\end{figure}

In Figure \ref{vfracs} we show the Tug-boat parameter $v_{\rm Tug}$ defined as the fraction between the velocity of the BH at the end of the simulation $v_{f}$ and the velocity expected from symmetric accretion $v_{\rm{sym}} = (M_{\rm bh, 2,t_0}/M_{\rm bh,2})v_{\rm kick}$, as a function of the natal kick for all the simulations. A parameter $v_{\rm Tug}>1$ indicates an efficient Tug-boat mechanism (asymmetric accretion flow) which leads to a larger kick velocity than expected from the same amount of mass accreted in a symmetric explosion. Meanwhile, a parameter $v_{\rm Tug}<1$ indicates that the asymmetric accretion flow, led to a smaller kick velocity than expected, reducing even more the final kick velocity.

As can be seen from Figure \ref{vfracs}, larger explosion energies mostly produce larger Tug-boat parameters, except for 4 simulations with the highest energies ($E_{\rm SN}=0.9E_{\rm bin}$) and large kicks ($\geq1000$ km/s) in stars with $M_{\rm preSN} \geq 21M_{\odot}$. We can also notice that small explosion energies ($E_{\rm SN} = 0.2E_{\rm bin}$) always lead to a negative $v_{\rm Tug}$ implying that the kick final velocity is further reduced by asymmetric accretion, reaching values as low as $\sim0.2$ in the most extreme scenario. 

For a given natal kick velocity, as the explosion energy increases, the $v_{\rm Tug}$ transitions from negative to positive values reaching up to $\sim 2$ times the velocity expected from symmetric accretion for the largest explosion energies. The energy at which this transition occurs seems to depend on the natal kick velocity and on the stellar model, making difficult to define a critical explosion energy or kick velocity at which it occurs. From Figure \ref{vfracs}, it seems that increasing the natal kick velocity not always leads to an increase in $v_{\rm Tug}$, and this behavior seems to also depend on the stellar structure.

What is interesting to note is that the most massive models (28$M_{\odot}$ and 34$M_{\odot}$), were the ones that obtained the largest $v_{\rm Tug}$ parameter, reaching almost 2 times the velocity expected from symmetrical accretion. Furthermore, the $13M_{\odot}$ stellar model obtained the smallest increase in $v_{\rm Tug}$ increasing the expected kick velocity only $\sim20\%$ by asymmetric accretion. As mentioned in \ref{AsymmetricICs}, for this same model and low explosion energies ($E_{\rm SN} =0.2E_{\rm bin}$ and $E_{\rm SN} =0.4E_{\rm bin}$ ), it was impossible to obtain kick velocities $\geq 1000$ km/s with our gaussian asymmetry prescription.


\subsection{Properties of the Binaries using the IML Approximation}

Using the isolated stars simulations, the binary properties from Table \ref{Tab:Bin}, and the formulation presented in Section \ref{IML}, we obtained the binaries' final orbital properties, assuming instantaneous mass loss. 

\begin{figure}
  \includegraphics[width=0.5\textwidth]{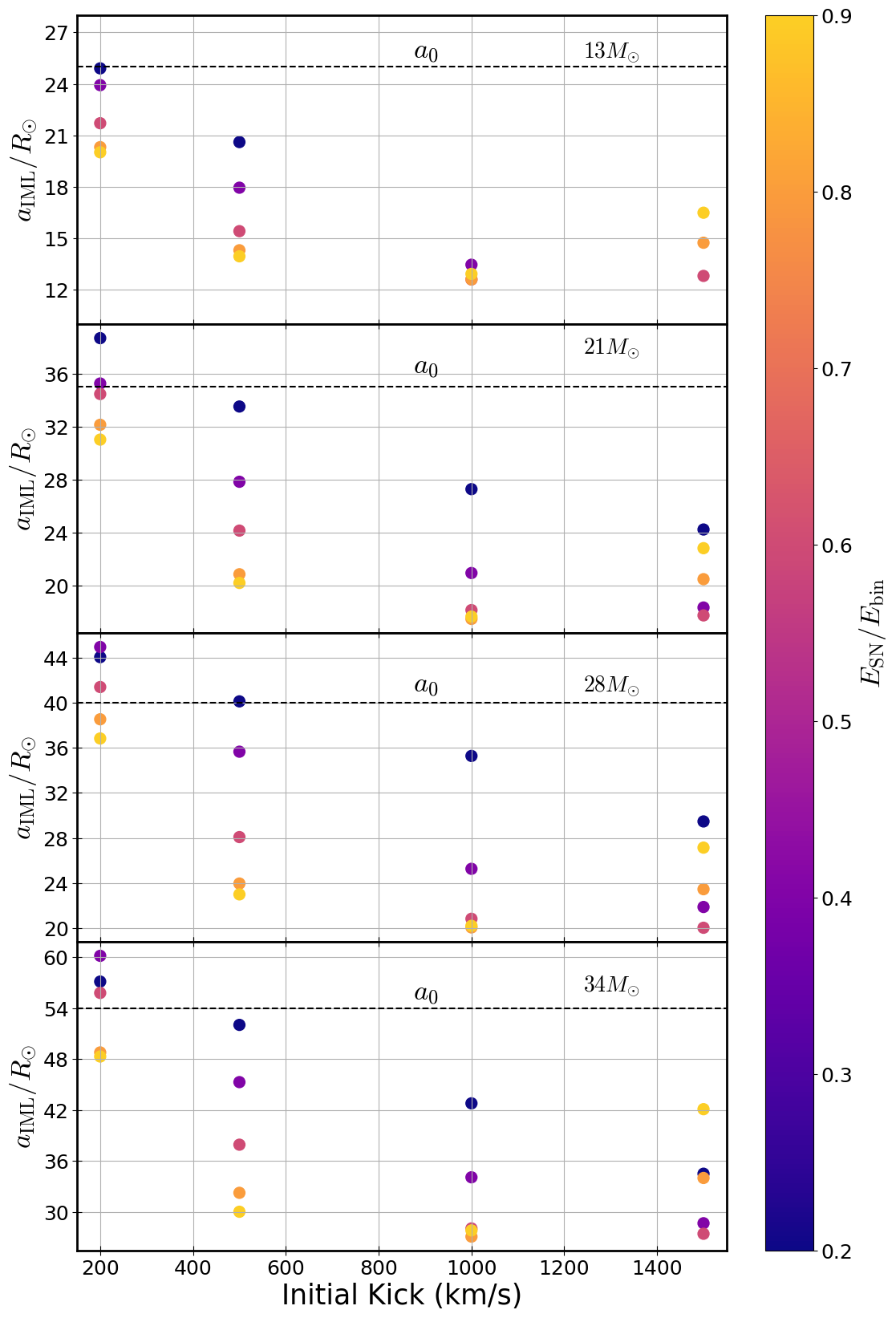}
  \caption{Final separation assuming instantaneous mass loss, vs BH natal kick, for all simulations. The dashed line indicates the initial orbital separation of the system $a_0$.}
  \label{Final_sepIML}
\end{figure}

Figure \ref{Final_sepIML} shows the final separation obtained for the binary systems defined in Table \ref{Tab:Bin} for the different explosion energies and kick velocities shown in Table \ref{Tab:SNEfracs}. As mentioned before, the direction of the kick is assumed to be in the orbital plane and opposite to the initial orbital velocity in order to study the best case scenario to allow the NK to produce a negative $\chi_{\text{eff}}$ without disrupting the binary. 
It is interesting to note that for all energies, the final separation decreases as the natal kick increases from 200 km/s to 1000 km/s, but when the natal kick is increased to 1500 km/s it leads to an increase in the final separation compared to the one obtained with a 1000 km/s kick for some explosion energies. 


\begin{figure}
  \includegraphics[width=0.5\textwidth]{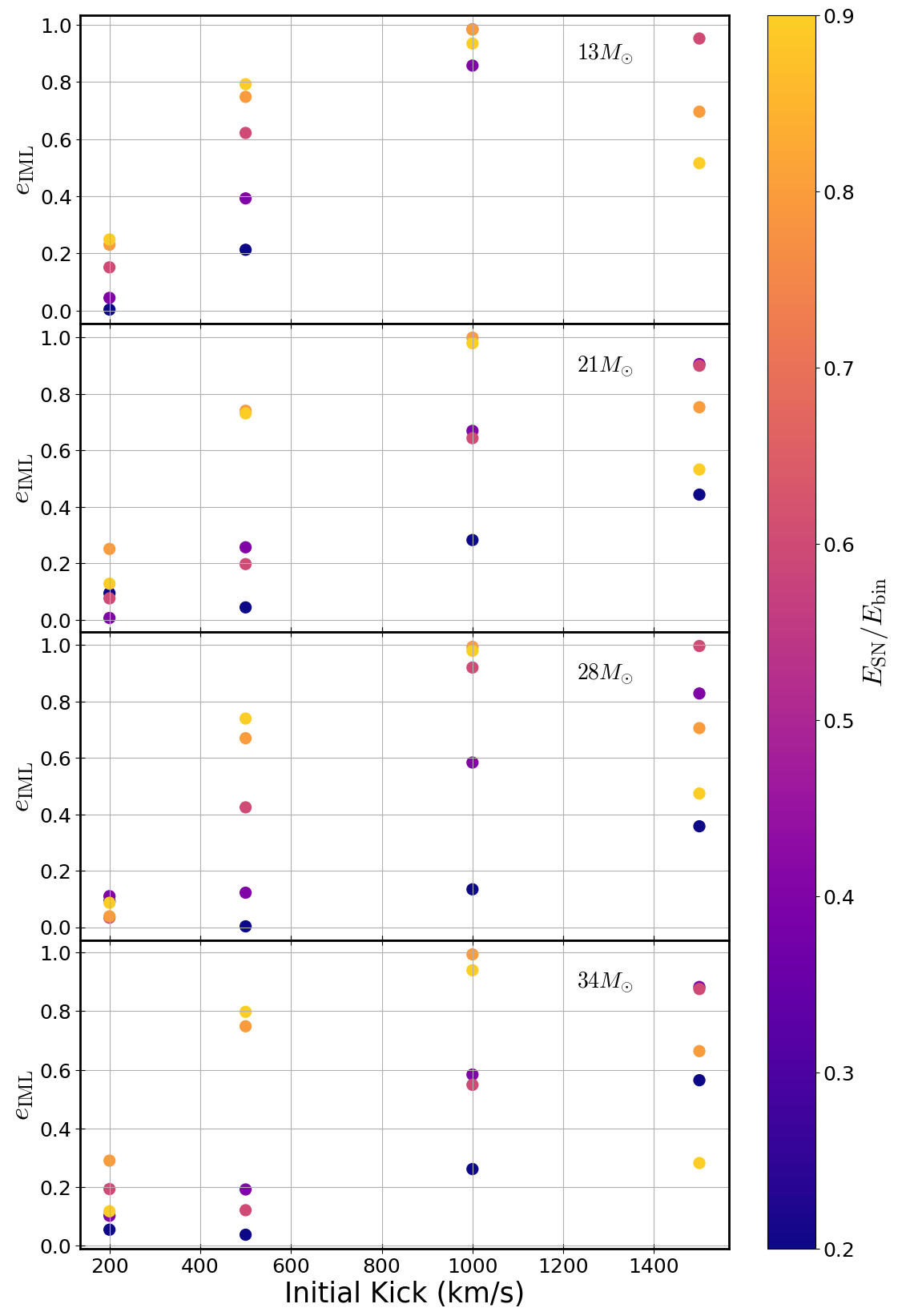}
  \caption{Final eccentricity (assuming instantaneous mass loss) vs BH natal kick, for different explosion energies and stellar masses.}
  \label{esIML}
\end{figure}

Another important result obtained from the IML approximation is the final eccentricity shown in Figure \ref{esIML}. As with the final separation, we also see a trend where increasing kicks up to 1000 km/s result in larger eccentricities for all energies, but the transition from 1000 km/s to 1500 km/s leads to a decrease in the eccentricity compared to the ones obtained for the 1000 km/s kick for some explosion energies. This is due to the fact that near the 1000 km/s kick, the final orbital speed is close to zero, and a higher kick increases the final orbital speed in the opposite direction, explaining the transition observed.

Even tough all resulting binaries are bound, some of them have eccentricities really close to $e=1$, which will be important when studying the coalescence or merger timescale of the systems. This timescale can be obtained from the orbital energy loss of the binary due to the emission of gravitational waves according to \cite{Peters_1964}:

\begin{equation}
    \begin{split}
    \frac{dE}{dt}&=-\frac{32G^4}{5c^5}\frac{M_{\rm bh 1} M_{\rm bh 2} M}{a^5} f(e), \ \text{where} \\
    f(e) &= \left(1+\frac{73}{24}e^2 + \frac{37}{96}e^{4}\right)(1-e^2)^{7/2},
    \end{split}
    \label{GWEnergyLoss}
\end{equation}

\begin{figure}
  \includegraphics[width=0.5\textwidth]{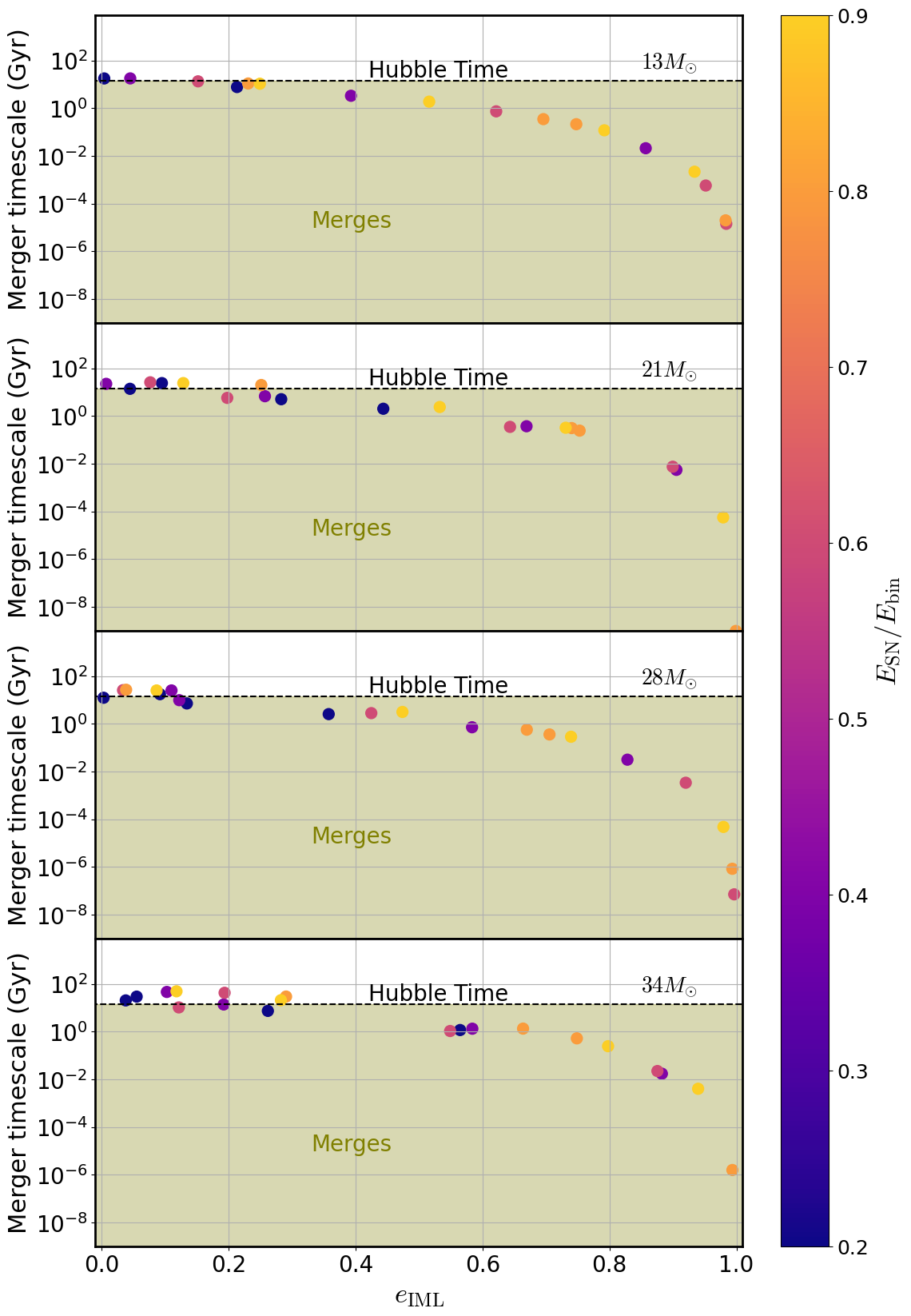}
  \caption{Merger timescales (assuming the final separation and eccentricity from IML), vs final eccentricity, for all the simulations with different explosion energies, BH natal kicks and stellar masses.}
  \label{tmerger}
\end{figure}

\noindent which depends on the individual masses of the components, their total mass $M$, the orbital separation $a$, and the eccentricity $e$. Solving equation \eqref{GWEnergyLoss} we can find the coalescence or merger timescale for all our binaries as shown in Figure \ref{tmerger}. As can be seen from the figure, the main driver of change in the merger timescale is the eccentricity, since the final separations shown in Figure \ref{Final_sepIML} are not so different from the original separation of the binaries, which lies lose to the maximum separation allowed for the original system to merge within a Hubble timescale. Furthermore, from Figure \ref{esIML} we can see that the largest eccentricities were mostly obtained for the largest explosion energies with natal kick velocities of 1000 km/s, but even systems with 500 km/s natal kicks are able to reach eccentricities above 0.5 which translate to merger timescales $\lesssim 0.1$ Gyrs according to the results shown in Figure \ref{tmerger}. 


\begin{figure}
  \includegraphics[width=0.5\textwidth]{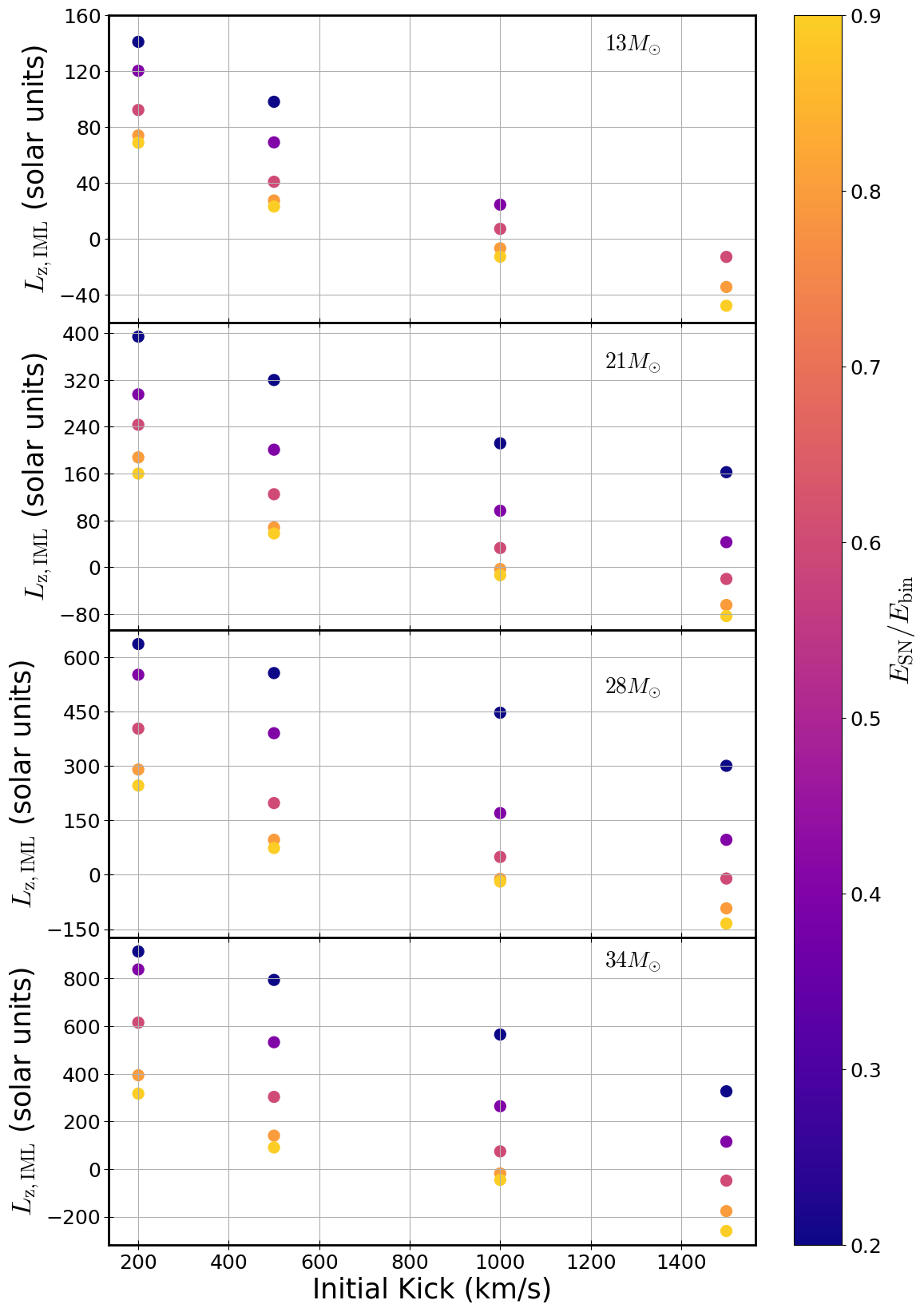}
  \caption{Final angular momentum $L_{z,\rm IML}$ assuming instantaneous mass loss, vs BH natal kick, for different explosion energies and stellar masses.}
  \label{Final_Lz}
\end{figure}

Finally, using equation \eqref{Lz}, we obtained the binaries' final orbital angular momentum $L_{z,\rm IML}$ shown in Figure \ref{Final_Lz}. As can be seen, increasing the explosion energy leads to a smaller $L_z$, explained partially by the increasing amount of mass ejected from the binary. Moreover, since the kick $v_{\rm kick}$ acts in the opposite direction of $v_{\rm orb}$, increasing the kick velocity also produces a reduction in $L_z$, which can ultimately lead to a change of direction of the angular momentum. However, as in the spherical accretion scenario represented by \eqref{vk_sym}, the initial kick velocity is significantly reduced by the accreted mass, thus, the actual kick that the BH receives after most of the accretion occurs (and the corresponding change in $L_z$) depends significantly on the fallback determined by the explosion energy.

As expected, in order to change the direction of $L_z$ the natal kick has to be considerably larger than the orbital velocity so that its reduction due to mass accretion still leads to a kick velocity larger than the original orbital velocity. For the systems explored in this work, this can only be achieved with the combination of high explosion energies ($E_{\rm SN}\geq0.6 E_{\rm bin}$) and large natal kicks ($v_{\rm kick}\geq$ 1000 km/s).

\subsection{Simulations of Exploding Stars in Binaries}
\label{binary}

In this section, we will study the simulations that include the binary companion BH, using a single natal kick of 1000 km/s and the rest of the properties shown in Tables \ref{Tab:SNEfracs} and \ref{Tab:Bin}. Given the results obtained from the instant mass loss approximation, we expect a natal kick of $1000$ km/s to produce some systems with negative $L_z$ (and $\chi{\rm eff}$) as in Figure \ref{Final_Lz}.

Since in these simulations the exploding star is not isolated, and the gravitational pull from the companion will affect the natal kick and the fallback of mildly ejected material, it is difficult to isolate the effect of the Tug-boat mechanism on the BH's velocity. Thus, we focused on the analysis of the binaries' properties, which can be compared to the IML approximation in order to discern the influence of the companion in the accretion process and the viability of using this approximation. 

\begin{figure}
  \centering
  \includegraphics[width=0.5\textwidth]{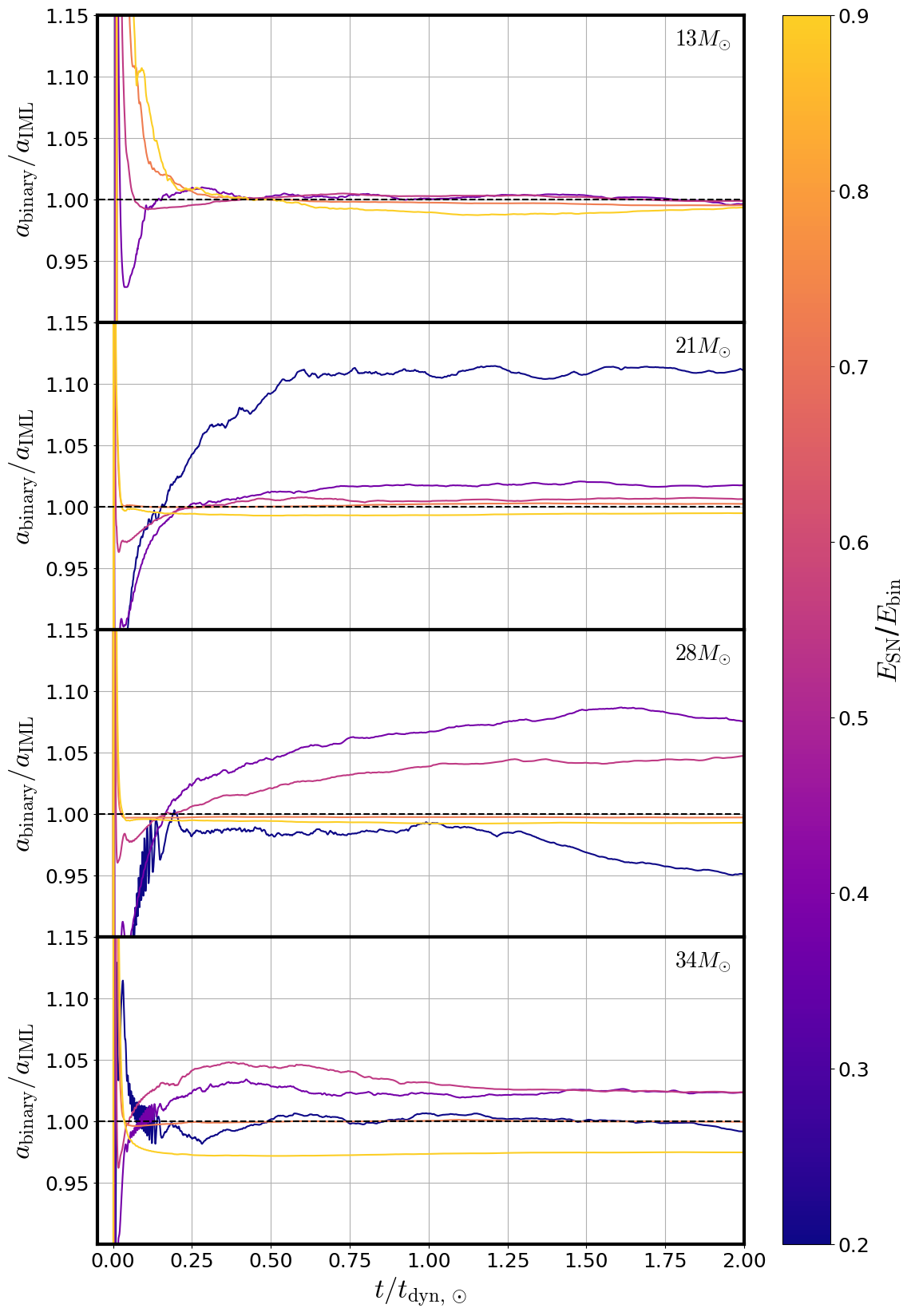}
  \caption{Final separation of the Binary system as a fraction of the calculation assuming instantaneous mass loss, vs time, for different explosion energies and stellar masses, a secondary star of mass $\approx 30/28$ times the initial star, and an initial BH velocity of $1 000$ km/s.}
  \label{afs}
\end{figure}

Figure \ref{afs} shows the evolution of the ratio between the final orbital separation obtained from the binary simulations $a_{\rm binary}$, and the one obtained from the IML approximation $a_{\rm IML}$. Thus, a value of $1$ implies that the orbital separation obtained from the binary simulation is the same as the ones shown in Figure \ref{esIML} for the 1000 km/s natal kick. As before, the color of the line indicates the explosion energy used, and each panel corresponds to a different exploding stellar model. 

As with the isolated star simulations, the result depends on the stellar model, but typically, simulations with the largest explosion energies (0.8 and $0.9E_{\rm SN}/E_{\rm bin}$) yield orbital separations similar to the IML approximation within less than 1\%. These large explosion energies provide larger radial velocity compared to low explosion energies. Since the orbital period is the characteristic timescale of each binary system, one expects that, if mass is removed much faster than this timescale, (as in the most energetic explosions) the result will resemble the IML approximation the most. However, as the time it takes to remove mass increases (for lower explosion energies), the orbital evolution of the system will increasingly differ from the IML approximation, even if the mass lost is the same. On the other hand, the $13M_{\odot}$ stellar model seems to have a peculiar structure that leads to a great agreement between the orbital separation obtained from binary simulations and the ones obtained from the IML approximation for all explosion energies. However, we see that the orbital separation results obtained from the binary simulations differ from the IML approximation by $\sim$12\% at most, as shown for the $21M_{\odot}$ stellar model. 


\begin{figure}
  \includegraphics[width=0.5\textwidth]{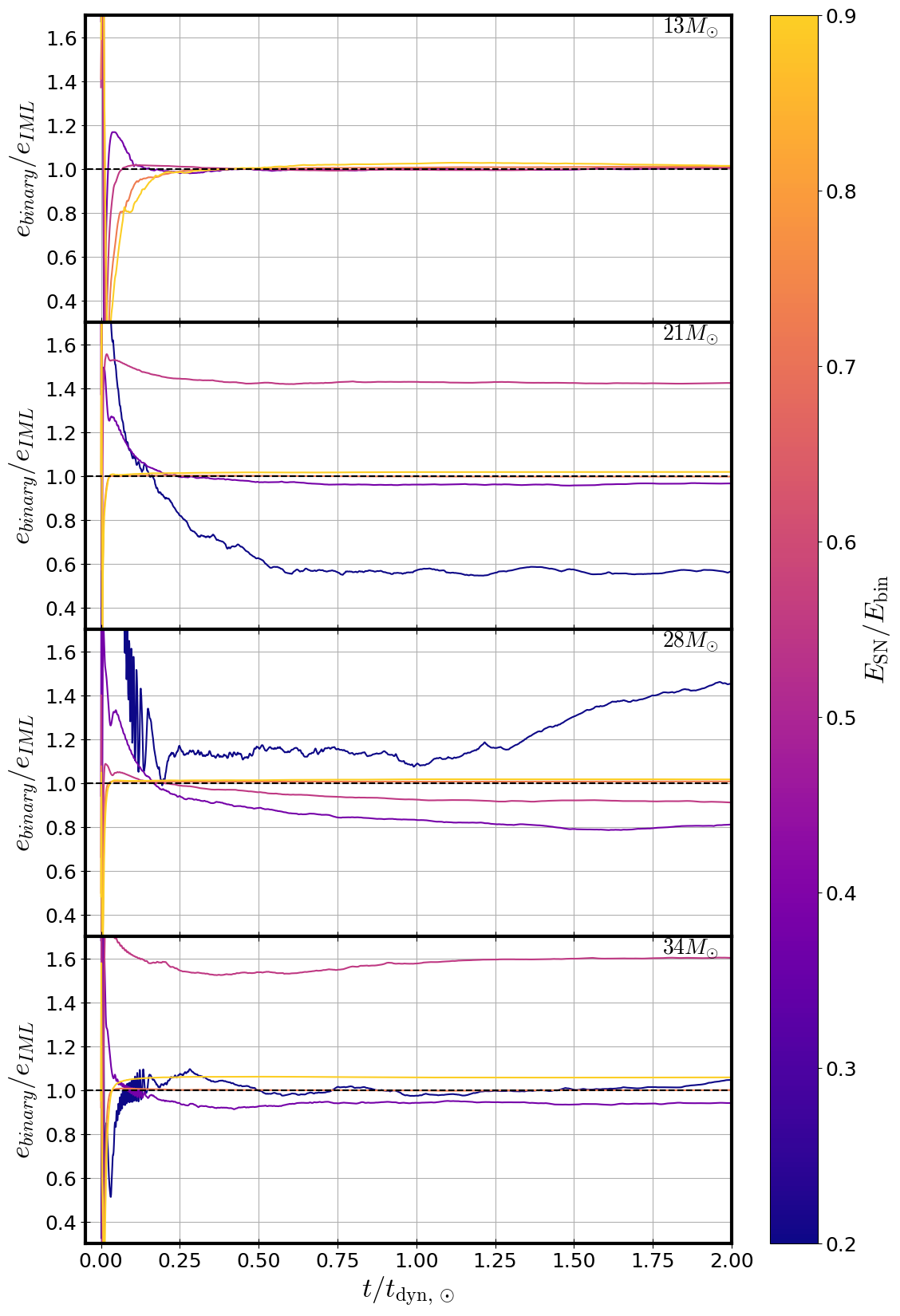}
  \caption{Final eccentricity of the Binary system as a fraction of the calculation assuming instantaneous mass loss, vs time, for different explosion energies and stellar masses, a secondary star of mass $\approx 30/28$ times the initial star, and an initial BH velocity of $1 000$ km/s.}
  \label{efs}
\end{figure}

Figure \ref{efs} shows the evolution of the ratio between the final eccentricity obtained from the binary simulations $e_{\rm binary}$, and the one obtained from the IML approximation $e_{\rm IML}$. As with the orbital separation, the $13M_{\odot}$ stellar model seems to have a peculiar structure that leads to a great agreement between results from binary simulations and the ones obtained with the IML approximation. As before, the higher explosion energies lead to eccentricities more similar to the ones obtained with the IML approximation. It is worthwhile to notice that the final eccentricity obtained from the binary simulations seems to be subject to larger variations than the orbital separation when compared with the results from the IML, leading to eccentricities that differ up to $\sim$60\% for some explosion energies. This could have significant implications when computing the merger timescale of the resulting BH binaries, which strongly depends on the orbital eccentricity.

\begin{figure}
\includegraphics[width=0.5\textwidth]{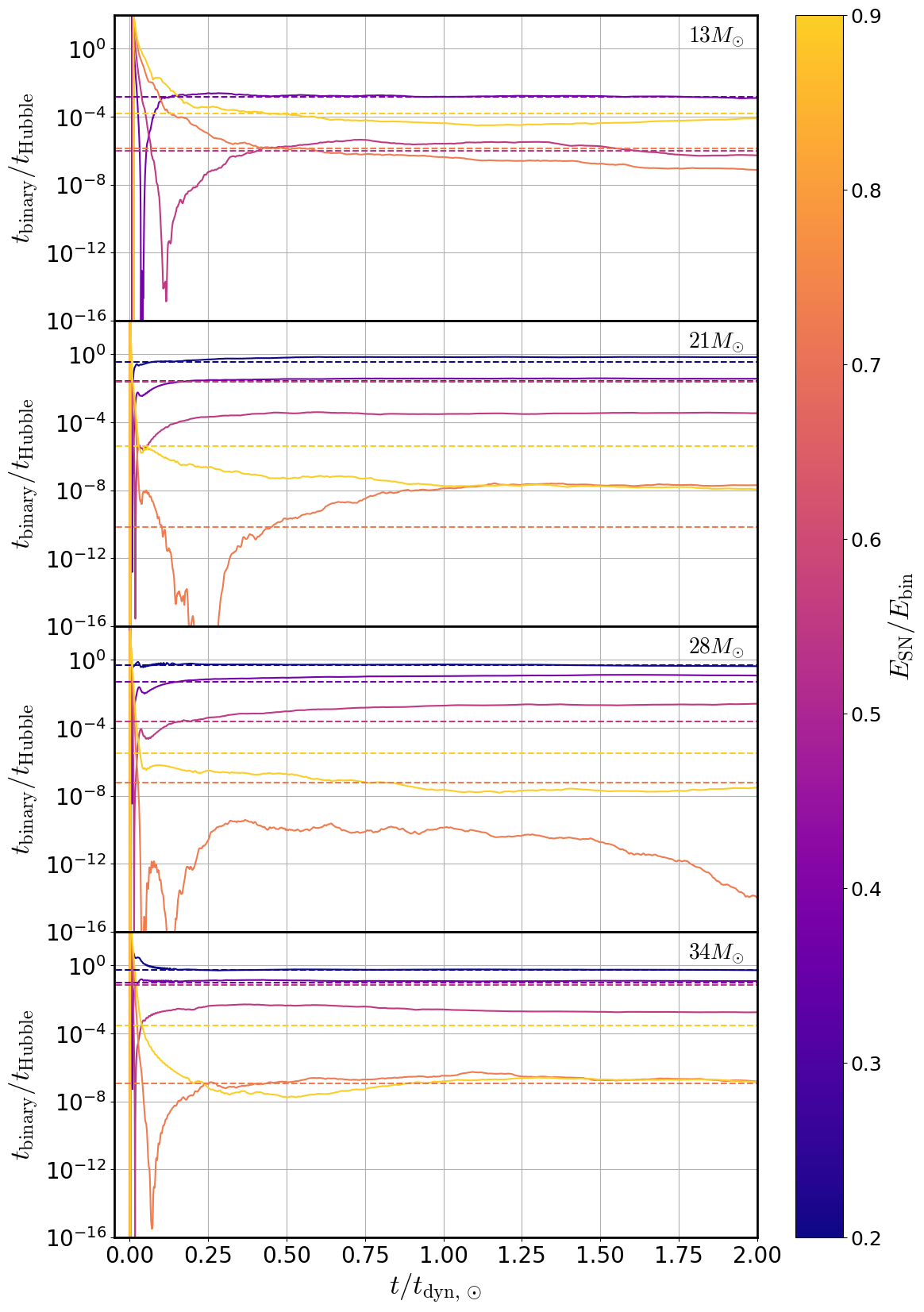}
  \caption{Calculated time of merger of the Binary system in proportion to the age of the Universe vs time, for different explosion energies and stellar masses, a secondary star of mass $\approx 30/28$ times the initial star, and an initial BH velocity of $1 000$ km/s.. The dashed horizontal lines represent the prediction made from the isolated simulations assuming instantaneous mass loss.}
  \label{tfs}
\end{figure}

As shown in Figure \ref{tmerger}, the merger timescale decreases with increasing eccentricity. Figure \ref{tfs} shows the evolution of the merger timescale obtained for the binary simulations. As before, each color corresponds to a different explosion energy and each panel to a different stellar model for the exploding star. The horizontal dashed lines indicate the merger timescale obtained for the IML approximation and the isolated exploding stars from Section \ref{isolated}. 

It is interesting to notice that higher explosion energies seem to lead to greater discrepancies from the IML approximation. This can be explained by taking into account that higher energies lead to higher eccentricities for the 1000 km/s natal kick, as shown in Figure \ref{esIML}. Moreover, since the change in the merger timescale as a function of eccentricity keeps increasing as $e$ approaches unity, the closer $e$ is to unity, a small variation in the eccentricity can yield orders of magnitude of difference in the merger timescale. 

\begin{figure}
  \includegraphics[width=0.5\textwidth]{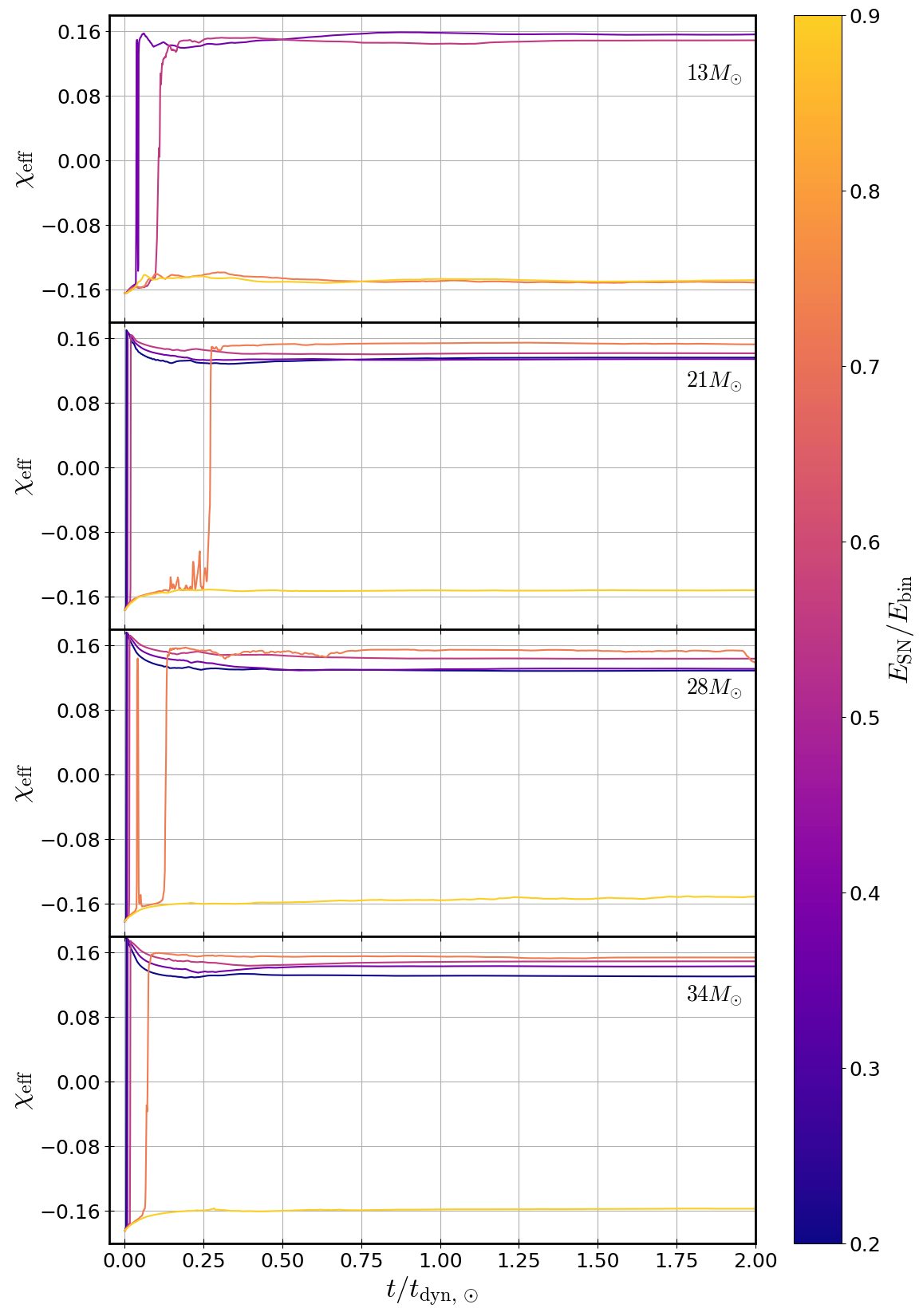}
  \caption{Effective spin parameter of the binary system vs time, for different explosion energies and stellar masses, a secondary star with an individual spin of $\chi_{2}=0.2$ and mass $\approx 30/28$ times the initial star, and an initial BH velocity of $1 000$ km/s.}
  \label{Xeffs}
\end{figure}

As mentioned in section \ref{initBinaries}, we assumed the primary component to have an initial spin of $S_{\rm BH, 1}\sim0.2$, considerably larger than the spin of the secondary ($S_{\rm BH, 2}\lesssim 0.06$), which makes easy to discern and associate any change in the sign of $\chi{\rm eff}$ with a change in sign on the orbital angular momentum. Figure \ref{Xeffs} shows the evolution of the effective spin parameter $\chi_{\rm eff}$ obtained from the binary simulations. As expected, the natal kick of 1000 km/s, being larger than the original orbital velocities shown in Table \ref{Tab:Bin}, leads to an initially negative effective spin parameter for all simulations. However, simulations with the most mass and momentum accreted (lower explosion energies) quickly reduce the BH's natal kick up to the point where further accretion of momentum forces the BH to move in the same direction as the preSN star, inverting the sign of $\chi_{\rm eff}$. This change in sign does not occur for simulations with the highest explosion energy $E_{\rm SN} = 0.9 E_{\rm bin}$, nor for the $13M_{\odot}$ preSN star with $E_{\rm SN} = 0.8 E_{\rm bin}$. Comparing these results with the ones obtained from the explosion of isolated stars and the IML approximation, we see that accounting for the gravitational influence of the binary companion, seems to complicate the production of binaries with negative effective spin, requiring a larger explosion energy for the BH to retain the direction of the original natal kick.

\begin{figure}
  \includegraphics[width=0.5\textwidth]{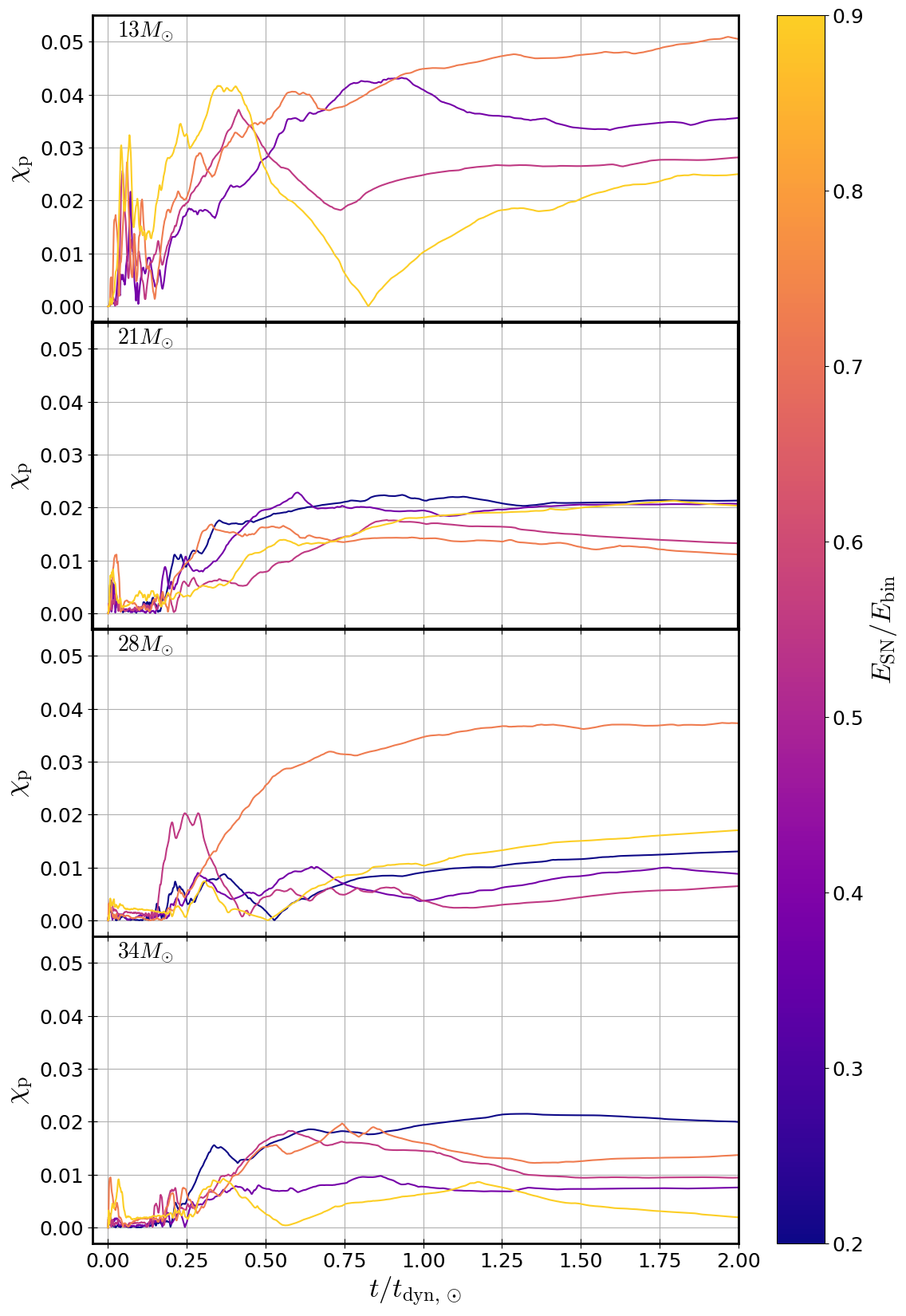}
  \caption{Precession spin parameter of the binary system vs time, for different explosion energies and stellar masses, a secondary star with an individual spin of $\chi_{2}=0.2$ and mass $\approx 30/28$ times the initial star, and an initial BH velocity of $1 000$ km/s.}
  \label{Precession}
\end{figure}

Another important piece of information that we can obtain from the BH binaries is the evolution of the components of the individual spins that lie on the orbital plane. This can be measured with the precession spin parameter $\chi_{\rm p}$, defined as:

\begin{equation}
 \chi_{\text{p}} = \frac{c}{B_{1}GM_{1}^2} \text{max}\left(B_{1}\boldsymbol{S}_{\rm bh, 1\parallel}, B_{2}\boldsymbol{S}_{\rm bh, 2\parallel}\right) > 0,
 \label{Xp}
\end{equation}

where, $\boldsymbol{S}_{\rm bh, i\parallel}$ is the projection of the individual BH spin on the orbital plane, $B_{1} = 2+3q/2$, $B_{2} = 2+3/(2q)$ and $q = M_1/M_2$. This parameter, together with the effective spin parameter, contains important information about the GW signal emitted by the binary during its orbital decay \citep{Abbott_2020}. The evolution of $\chi_{\rm p}$ is shown in Figure \ref{Precession}, and as can be seen, there is a non-zero spin along the orbital plane obtained by the secondary BH as it accretes the envelope. Part of this non-zero spin on the orbital plane is explained by the asymmetric distribution of particles in the initial conditions, in which the rotation and radial velocities produce an asymmetry in the angular momentum components along the $x$ and $y$ axis. Nevertheless, in all simulations, the angular momentum components along the orbital planes ($L_{x,\rm PreSN,}$ and  $L_{y,\rm PreSN,}$) are over 2 orders of magnitude smaller than the angular momentum $L_{z,\rm PreSN,}$ determined by the angular velocity given to all stellar models.


\section{Discussion}


\subsection{On the asymmetry of the explosion and the Tug-Boat mechanism}

Given the results showed in Section \ref{isolated}, it is clear that the Tug-boat mechanism (the asymmetric accretion of material) played an important role in determining the final velocity of the BH. However, as seen in Figure \ref{vfracs}, the resulting kick is conditioned by the explosion energy, the intensity of the asymmetry (the natal kick) and the specific mass distribution of the exploding star. Results show that the largest explosion energies ($E_{\rm SN}/E_{\rm  bin}\geq 0.8$) consistently lead to an increase in the expected natal kick from symmetric accretion ($v_{\rm Tug}>1$), meanwhile, the lowest explosion energy ($E_{\rm SN}/E_{\rm  bin}\leq 0.2$) always leads to a decrease in the expected natal kick ($v_{\rm Tug}<1$).

We should take into account that this work only explored a single gaussian width $\sigma=0.6$ which, along with the parameter $\Delta$, determine the shape and size of the asymmetry. However, our setup may be completely different from the asymmetries obtained from neutrino emission or other physical mechanisms that produce the mild explosion. Thus, our results should not be taken as a generalization, but as evidence that for certain type of asymmetries, the Tug-boat mechanism can lead to an increase or a decrease in the expected natal kick when compared with the one obtained from symmetric accretion of material. 

However, the intensity and shape of the asymmetries determined by the parameters $\Delta \in [0.25,400]$ and $\sigma=0.6$ must be related to the momentum  asymmetry parameter $\alpha_{\rm ej}$, defined in \cite{janka2017} as the ratio between the magnitude of the integrated linear momentum of the gas surrounding the compact object $\mathbf{P}_{\rm gas}=\int \rho \mathbf{v}\ dV$ and the momentum this material would have if it moved in a single direction $P_{\rm ej}=\int \rho |\mathbf{v}|\ dV$. In our scenario,  the asymmetry of the explosion only appears towards $\theta=0$, and the velocity that matters inside the integral determining $\mathbf{P}_{\rm gas}$ is the projection of the radial velocity onto the $z$ axis, $\mathbf{v}=v_r\cos \theta\  \hat{\mathbf{k}}$, while the other components cancel each other out in the integration process. Thus, by obtaining the radial velocity magnitude from \eqref{eSN_part} assuming a continuous mass distribution (i.e. $m_i\rightarrow dm$ and $\theta_i = \theta$), the momentum  asymmetry parameter can be written as:
\begin{equation}
    \alpha_{\rm ej}=\frac{|\mathbf{P}_{\rm gas}|}{P_{\rm ej}} = \frac{\int_0^{\pi} \left[f(\theta,\Delta,\sigma) \right]^{1/2}\cos{\theta}\sin{\theta}\ d\theta}{\int_0^{\pi} \left[f(\theta,\Delta,\sigma) \right]^{1/2}\sin{\theta }\ d\theta}.
\end{equation}

For the parameters $\Delta$ and $\sigma$ used in this paper, $\alpha_{\rm ej}$ ranges from $0.013$ to $0.37$ except in one scenario, which is consistent with previous results in the literature  \citep[e.g. ][]{Sheck2006}. Only the simulation with preSN mass of $13M_{\odot}$ and $E_{\rm SN}/E_{\rm bin}=0.6$ required a larger value of $\alpha_{\rm ej}=0.49$ (corresponding to $\Delta \sim 400$) in order to obtain the most extreme natal kick used in our simulations ($v_{\rm kick}= 1500$km/s). And as mentioned before, for the same stellar model and smaller explosion energies, it was impossible to obtain kick velocities $\geq 1000$ km/s with our gaussian asymmetry prescription.

Figure \ref{vfracs} shows that the asymmetric explosion and the subsequent accretion of mass and momentum lead to a kick velocity that can differ significantly from the one expected after symmetric accretion $v_{\rm sym}$. This shows that the Tug-boat mechanism took place in our simulations, and that its effect on the natal kick is more complicated than expected, since it can yield either a reduction or an increase in the natal kick after accretion. However, it seems that as the explosion energy approaches the binding energy of the envelope it is more likely to obtain larger natal kicks than the ones expected from symmetric accretion. Meanwhile, explosion energies below $0.4E_{\rm bin}$ tend to lead smaller natal kicks than the ones obtained with symmetric accretion. Interestingly, even though the momentum asymmetry parameter $\alpha_{\rm ej}$ (determined by our parameters $\Delta $ and $\sigma$) increases with the initial natal kick, it seems to have little effect on the parameter $v_{\rm Tug}$, and the explosion energy and the fallback it produces seems to be the main driver of this parameter.


\subsection{On the amount of fallback}

\begin{figure}
  \includegraphics[width=0.5\textwidth]{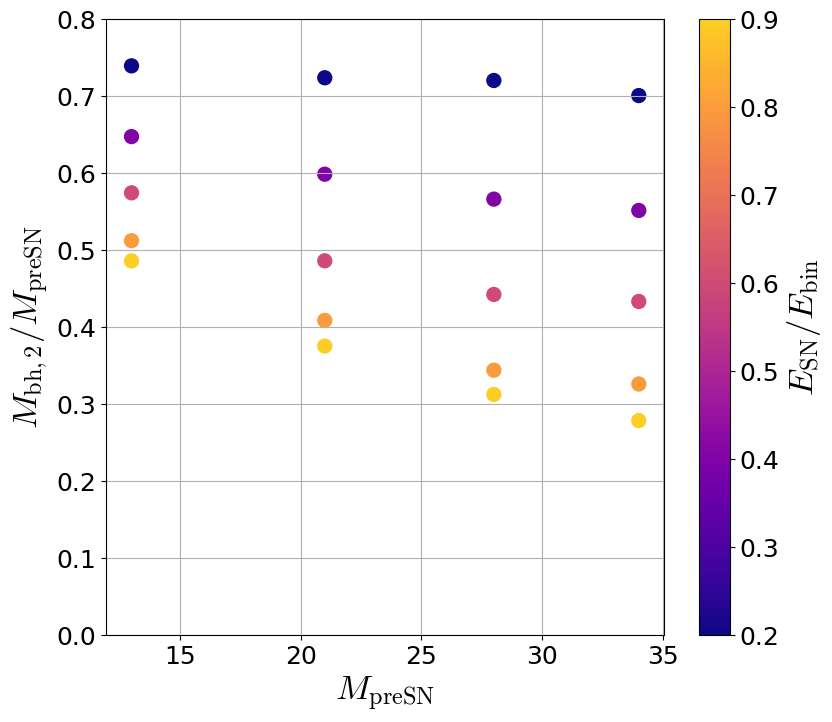}
  \caption{Final fallback, defined as the final mass of the BH in proportion to the preSN mass, vs the preSN stellar mass, for different explosion energies, averaged over the different NK speeds. The colorbar represents the energy of the explosion, as a percentage of the gravitational binding energy ($E_{\rm SN}/E_{\rm bin}$). The specific values utilized are in Table. \ref{Tab:SNEfracs}.}
  \label{Fallback}
\end{figure}

Figure \ref{Fallback} shows the resulting fallback, defined as the ratio between the final BH mass and the initial preSN stellar mass, as a function of preSN stellar mass, for different SN explosion energies, and averaged over the different NKs. As expected, the main driver of the amount of fallback is the explosion energy, with higher energies leading to larger mass loss and reduced fallback. Interestingly, for a given explosion energy, there seems to be a clear downward trend between fallback and increasing the preSN stellar mass.

\begin{figure}
    \centering
    \includegraphics[trim={1.5cm 0.5cm 2.5cm 1.5cm}, width=0.95\linewidth]{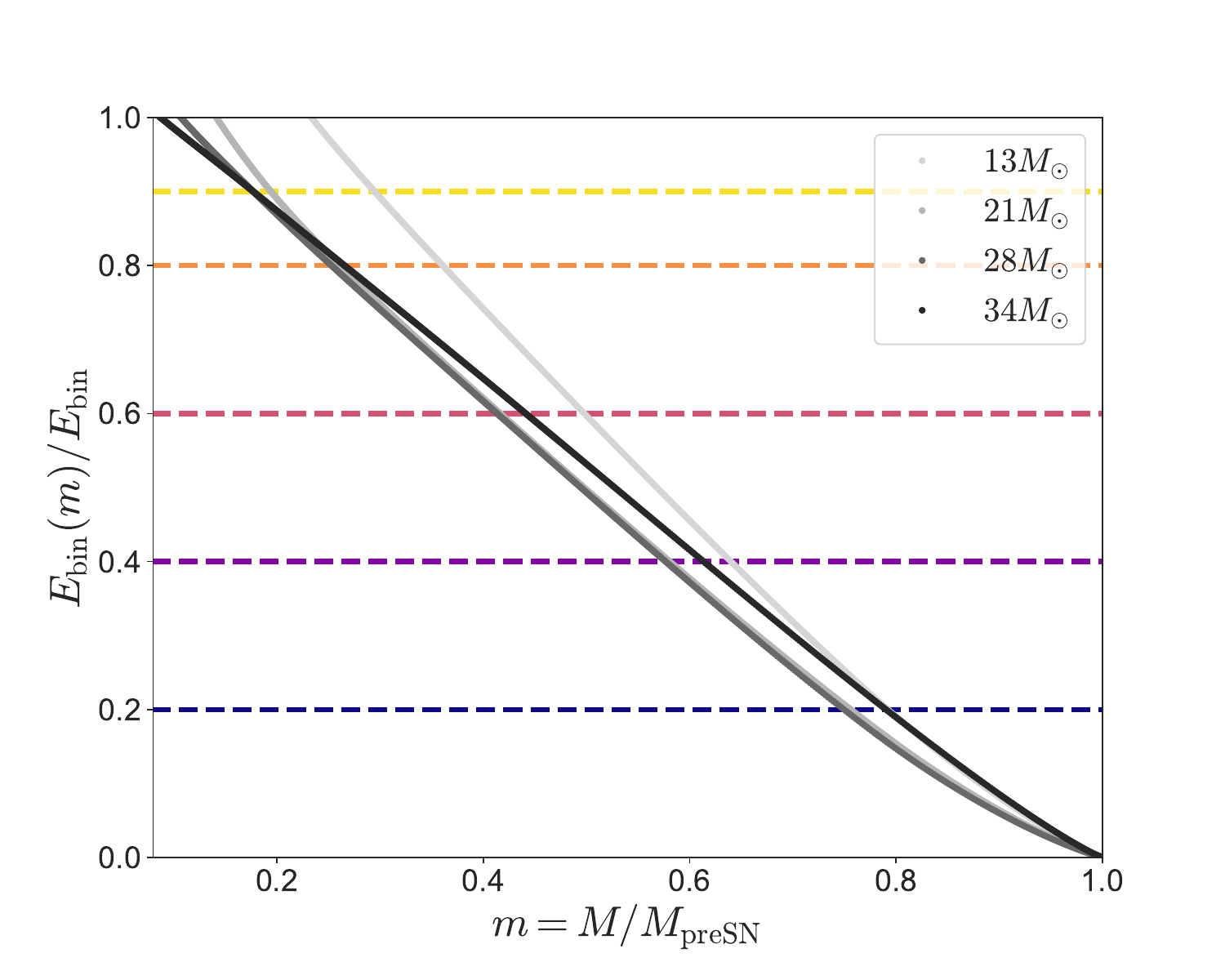}
    \caption{Binding energy $E_{\rm bn}(m)$, normalized by the binding energy of the envelope surrounding the black hole ($E_{\rm bn}$) as a function of the normalized interior mass of each star, $m = M(r)/M_{\rm PreSN}$. The horizontal dotted lines represent the explosion energies used in our simulations. The intersection of the explosion energies and the binding energy profiles provides an estimation of the fallback expected for each scenario and star.}
    \label{BinEn}
\end{figure}

This trend can be better understood with Figure \ref{BinEn}, which shows the binding energy $E_{\rm bn}(m)$ profile of the four stellar models (grey lines), normalized by the binding energy of the envelope surrounding the black hole $E_{\rm bn}$, as a function of the normalized interior mass of each star $m=M/M_{\rm presN}$.  By definition the grey curves reach $E_{bn}(m)/E_{bn} = 1$ precisely at $m = 3 M_{\odot}$ (the mass of the BH). The horizontal dotted lines represent the explosion energies used in our simulations. 

We can infer a few things from this figure. Given an explosion energy $E_{\rm SN}$, this energy should be sufficient to unbind only those parts of the star for which $E_{\rm bn}(m) < E_{\rm SN}$. Therefore, the points where the horizontal lines intersect the curves for each star provide an estimate of the amount of mass that could be ejected—or, conversely, the amount that might fall back. From this analysis, one would be able to explain that for lower energies, the fallback are similar between them, and as energy increases, the $13M_{\odot}$ star has a significantly higher fallback than the other systems. The most relevant factor here would be how compactly the material is distributed in the initial conditions. 

However, this approach does not fully explain Figure \ref{Fallback}, since the fallback from the $34M_{\odot}$ model is the lowest for any given explosion energy, and this behavior is not shown in Figure \ref{BinEn}. This disagreement can be cause for several reasons. Firstly, the biding energy $E_{\rm bn}$ does not account for the thermal energy of the material, which would render it less "bound" than the $E_{\rm bn}$ values suggest. Consequently, it would not be unusual to observe less fallback than what is indicated in Figure \ref{BinEn}. Furthermore, this does not show how much of the explosion energy is dissipated as heat during the process; which could alter the amount of material actually ejected—or determine whether the material is merely heated while remaining gravitationally bound. Thus, while this graph provides a general indication of the expected fallback, the precise value would likely also depend on the star's internal energy distribution and the specific manner in which the shock wave transferred its energy. The results shown in Figure \ref{Fallback} stem from a complex interplay between the explosion energy, the internal mass profile of each preSN stellar model, the asymmetry and how the material is distributed afterwards, which makes a single phenomenon unlikely to fully explain this results.

\subsection{Negative effective spin parameters from natal kicks}

The pattern from Figures \ref{Final_Lz} and \ref{Xeffs} is much clearer. Because the initial conditions were programmed so that the NKs act perfectly against the orbit, higher explosion energies yields lower values for $L_z$. This is unsurprising. However, our simulations found that, contrary to what was predicted in Fig. \ref{Final_Lz}, a NK of $\sim 1 000$ km/s with energy of $0.8E_{\rm bin}$ may not change the direction of the orbit. Instead, this combination of NK and energy only changes the direction of the orbit for the $13M_{\odot}$ system. We emphasize that this is important because it is a prediction that is assuming neither spherically asymmetric accretion nor IML. It is also worth noting that NKs of $\sim 1000$km/s are amongst the highest observed for neutron stars, and even under these circumstances it may not be able to invert the orbit. This implies that there may be other mechanisms at play to explain negative effective spin parameters, like the ones mentioned in \cite{baibhav}. 

This analysis is done with one of the highest NKs observed for neutron stars, and even in this particular case, the NK did not consistently change the direction of the orbit.
An important caveat in this analysis is the orbital properties of the binary systems we began with (Table \ref{Tab:Bin}). The reason the systems were compact to begin with, is that normally one expects these systems to be close enough so that they can merge before the Hubble time. However, taking a look at the merger timescales after the NK (Fig. \ref{tmerger}), it is possible for the NK to reduce significantly the merger timescale by making the orbit more eccentric. This result is also consistent with the second figure in \cite{Vigna2025}, where he shows that tangential kicks opposite to the orbit can significantly decrease the merger timescale, as opposed to radial or polar kicks. Therefore, in theory, one could begin with a wider binary that would not merge before the Hubble time under normal circumstances, and a more realistic NK may produce an orbit eccentric enough for the system to have negative spin and also merge before the Hubble time. This is something that will be further analyzed in a future paper.

\subsection{On the IML approximation}

To understand under which circunstances IML is a reasonable approximation, we can take a look at Figs  \ref{esIML}, \ref{afIML}, \ref{afs} and \ref{efs}. As mentioned in the previous section, there is a significant discrepancy in explosions with lower energies between the actual accretion we found, and the ones predicted by assuming IML. However, both in final separation and in final eccentricity, this discrepancy appears to also vary from one stellar mass to the next. It seems that for all stellar masses, SNE energies of $0.8$ and $0.9$ times the gravitational binding energy of the star, which in turn corresponds to a faster SNE, is consistent with the IML approximation. However, for lower energies the result could be significantly different from the approximation. The exception to the rule is the $13M_{\odot}$ star. For these simulations, the IML approximation seems to apply better than for the other stellar masses. It is unclear why there does not seem to be a pattern in which simulations will be more consistent with the IML approximation, and more research is needed on this topic.

\section{Conclusions}

In this paper we addressed 2 main questions: whether the gravitational tug-boat mechanism takes place in our asymmetric explosions and whether the resulting NKs are able to change the direction of the orbit and the sign of $\chi_{\text{eff}}$.

It has been proposed that a BH NK of linear momentum can be theoretically able to change the direction of the orbit, and as a result, the sign of $\chi_{\text{eff}}$ in BHBs. This NK can be the result of an asymmetric SNE \citep{janka2013, brandt}. Theoretical calculations have been made to determine how should the distribution of BH NKs be, and how big the NK should be to change the sign of $\chi_{\text{eff}}$. In particular, \cite{janka2013} proposed that the BH NKs should be bigger than the reduced NKs of Neutron Stars after a spherically symmetric accretion of mass and momentum.

In order to study this scenario we analyzed the process of accretion of material by a newly formed BH located in the center of an exploding star, whose NK will be determined by the asymmetry of the explosion. This was done by means of 3D SPH simulations of pre-supernova stellar models with $13{M}_{\odot}$, $21{M}_{\odot}$, $28{M}_{\odot}$ and $34{M}_{\odot}$ from \cite{heger}. This allowed us to follow the process of accretion, and the entire evolution of the BH's properties (including the NK). 



A wide range of NKs was simulated by changing the amount of asymmetry in the explosion. There was also variation in the SNE energies, given as percentages of the gravitational binding energy of the star. We calculated the hypothetical final separation and eccentricity of a system comprising our simulated star and a companion with slightly higher mass, and a separation in the limit of what would be observable with GWs. Additionally, the time of merger and maximum final separations allowed for the systems to be observable were analyzed. 

After running these simulations and analyzing the results, additional simulations of a binary system were run with $8\times 10^{6}$ SPH particles, for all stellar masses. The details of these runs are found in Table \ref{Tab:Bin}. We calculated the separation and eccentricity of the system and compared them with the ones obtained from the IML approximation. We also calculated the orbital angular momentum measured from the center of mass and the effective and precession spin parameters of the binary. 

Our conclusions are as follows:
\begin{enumerate}
    \item Higher explosion energies will result in BBHs with unequal masses. Most observations are consistent with mass fractions closer to 1, but systems with unequal masses have been observed before.
    \item The final velocity of the BH after accretion can be lower, equal, or larger than expected from spherically symmetric accretion ($v_{\text{sym}}$), depending on the energy of the explosion. SNEs with an energy of 20\% of the gravitational binding energy of the star gave consistently lower values than $v_{\text{sym}}$, whereas SNEs with a percentage of 60\% or more gave final velocities higher than $v_{\text{sym}}$. This indicates that the mechanism described by \cite{janka2013} may be at play.
    \item Whether the gravitational tug-boat mechanism does happen or not, it is undeniable that the asymmetrical accretion must be taken into account because it significantly alters the result of the final velocity of the BH after accretion. The amount of the alteration depends primarily on the SNE energy and mildly on the NK. 
    \item The asymmetries resulting from the setup presented in this work are equivalent to values of the asymmetry parameter from \cite{janka2017} of $\alpha_{\rm ej}\in[0.013,0.37]$  in all cases but one, which is consistent with \cite{Sheck2006}. 
    \item We find a nontrivial relation between the amount of fallback after accretion and the preSN stellar mass. This is most likely a complex interplay between the asymmetry, the internal mass profile in the initial conditions, and the evolution of the system during the explosion.
    \item The final separation and eccentricity of these systems with high velocity NKs are indeed observable by GWs, mainly due to the fact that they end up with eccentricities that help them merge before the Hubble time. Some of our simulations with NKs of 200km/s had final separations larger than the maximum allowed for them to be observable, but this are not the types of systems that would change the orientation of the orbit regardless.
    \item The times of merger are very sensible to eccentricity when the kick is tangential to the orbit, which means that one must be careful of taking conclusions from the times of merger calculated using the IML approximation. 
    \item Even NKs of $\sim 1000$km/s, which are among the highest velocities observed for NS, were not able to consistently change the orientation of the orbit, thus returning a negative value of $\chi_{\text{eff}}$. The best case scenario for negative $\chi_{\text{eff}}$, which are systems with high observable initial separations, and a kick perfectly directed against the orbital velocities, only change the orientation of the orbit in the simulation with SNE energy of either 80\% or 90\% the gravitational binding energy of the star, depending on the stellar mass. 
    \item The final separation of the binary systems changes by no more than 10\% from the calculations done by assuming an IML in the SNE. However, the eccentricities can change dramatically if the IML approximation is not assumed. Explosions with energies of around 80\% the gravitational binding energy of the star or higher gave results very similar to the calculations, which means that the IML is a good approximation in those cases.
    \item The precession spin parameter changes during the accretion process without following a specific pattern and does not increase more than $\chi_{\rm p}=0.05$.
\end{enumerate}

\section*{Acknowledgements}

The authors thank Alejandro Vigna-Gómez for their useful commentary on the work presented. The authors acknowledge the National Supercomputing Laboratory of Southeastern Mexico (LNS) at BUAP for the computational support and supercomputing resources provided for this work, under project 202004059C. ESL y CAN acknowledge support from the SECIHTI graduate fellowship. AB acknowledges support from the IIxM program by SECIHTI.

\section*{Data Availability}

 All data and codes are available upon request to the authors.



\bibliographystyle{mnras}
\bibliography{example} 


\bsp	
\label{lastpage}
\end{document}